\documentclass[a4paper, 12pt]{article}
\usepackage{mlmodern}
\usepackage{microtype}
\usepackage{amsmath, amssymb}
\usepackage{bm}
\usepackage[dvipsnames]{xcolor}
\usepackage{cite}
\usepackage{titling}
\usepackage{authblk}

\usepackage{tikz}
	\usetikzlibrary{calc}
	\usetikzlibrary{arrows.meta}
	\usetikzlibrary{decorations.markings}

\usepackage[
    colorlinks,
    citecolor = Green,
    linkcolor = blue,
    urlcolor  = blue
]{hyperref}

\usepackage[
    left   = 2cm,
    right  = 2cm,
    top    = 2.5cm,
    bottom = 2cm
]{geometry}

\begin{document}

\title{The large $|U|$ expansion for a half-filled asymmetric Hubbard model on a triangular ladder in the presence of spin-dependent magnetic flux}

\author[a,b,\thanks{\href{mailto:shota_garuchava@hotmail.com}{shota\_garuchava@hotmail.com}}]{{\normalsize Shota Garuchava}}

\author[a,b]{{\normalsize Bachana Beradze}}

\author[c]{{\normalsize Tatia Sharia}}

\author[a,b]{{\normalsize George I. Japaridze}}

\affil[a]{\small\textit{Ilia State University, Cholokashvili Avenue 3-5, 0162 Tbilisi, Georgia}}

\affil[b]{\small\textit{Andronikashvili Institute of Physics, Tamarashvili str.~6, 0177 Tbilisi, Georgia}}

\affil[c]{\small\textit{M. Nodia Institute of Geophysics, Alexidze str.~1, 0171 Tbilisi, Georgia}}

\date{}

\maketitle

\renewcommand{\abstractname}{\vspace{-5\baselineskip}}
\begin{abstract}

We consider a half-filled system of spin-1/2 fermions on a triangular ladder with spin-dependent hopping in the presence of spin-dependent flux. Using the Schrieffer-Wolff transformation, we derive an effective spin Hamiltonian describing the infrared properties of the system in the limit of strong on-site repulsion. The expansion is performed up to the third order in the hopping amplitude over on-site interaction energy. It is shown that the low-energy spin excitations are described by an anisotropic $XXZ$ Heisenberg model with Dzyaloshinskii-Moriya interaction and an unconventional three-spin (correlated exchange) term in the presence of an extended magnetic field. The latter couples not only to $z$-projection of the total spin on a triangular plaquette but also to the product of $z$-components of three spins located on vertices of the triangle. In the case of strong attractive interaction, the corresponding effective model is derived from the repulsive one using the one-spin-component particle-hole transformation. The effective Hamiltonian is given by the same $XXZ$ Heisenberg model, as in the repulsive case, but in terms of pseudospin (charge) operators.

\end{abstract}


\section{Introduction}
\label{intro}
The idea that the lattice models with more than two-body interactions may lead to a rich variety of critical behavior found its confirmation in several two-dimensional (2D) models of statistical physics:
the Ashkin-Teller model~\cite{AT_1943},
the Baxter model~\cite{Baxter_1972} and the Baxter-Wu model~\cite{Baxter_Wu_1973} involving four- and three-spin interactions, respectively. Their critical behavior differs drastically from that of Onsager's solution of the 2D Ising model. Later, several interesting models with multi-spin interaction have been proposed and studied~\cite{Gross_Mezard_84, Gardner_85, m-spin_Ising_Kolb_Penson_86, 3-spin_1988_Penson, m-spin_Ising_1998, 3-spin_Ising_1999, 3-spin_Ising_2009} mostly in the context of spin glass physics~\cite{Gross_Mezard_84,Gardner_85} and quantum statistics~\cite{m-spin_Ising_Kolb_Penson_86, 3-spin_1988_Penson, m-spin_Ising_1998, 3-spin_Ising_1999, 3-spin_Ising_2009}.

Understanding the ground states of frustrated quantum spin systems -- in which the local energetic constraints cannot all be simultaneously satisfied is a fascinating topic in condensed matter physics \cite{Balents_Nature_2010}. The Heisenberg antiferromagnet on a 2D triangular lattice is a paradigmatic model to study the effects that emerge due to the frustration. The Anderson's resonating valence bond state~\cite{Anderson_RVB_87}, a collective spin singlet not breaking any symmetry and possessing neutral spin-1/2 excitations, is a prime example that opened the way for topological phases with fractionalized excitations emerging in frustrated spin systems~\cite{TP_w_FrcExc_1, TP_w_FrcExc_2, TP_w_FrcExc_3, TP_w_FrcExc_4, TP_w_FrcExc_5, TP_w_FrcExc_6}. In 1987, 
Kalmeyer and Laughlin~\cite{Kalmeyer_Laughlin_87} proposed the chiral spin liquid (CSL) state that breaks time-reversal and parity symmetries as a different spin singlet state which is realized in the ground state of the spin ${S=1/2}$ triangular Heisenberg antiferromagnet. Soon afterward, Wen, Wilczek and Zee~\cite{WWZ_89} and also Baskaran~\cite{Baskaran_89}, proposed to use the expectation value of the scalar spin chirality operator 
\begin{align}
	\chi_{ijk}
	\sim
	\bm{S}_i
	\cdot
	\left[\bm{S}_j \times \bm{S}_k\right]
,
\end{align}
where $i, j, k$ belong to an elementary triangle, as an order parameter for CSLs. These observations boosted further activity in studies of spin systems with multi-spin coupling. It has soon been realized that the models with chiral three-spin coupling naturally emerge in the case of the half-filled Hubbard model on a triangular lattice in the presence of a magnetic field and strong on-site repulsion \cite{Sen-Chjtra_95}.

The effects caused by an external magnetic field that quantizes the spectrum of charged carriers in crystalline lattices has long been one of the fundamental problems in condensed matter physics \cite{Azbel_64, Brown_64, Hofstadter_76}. The kinetic frustration caused by the magnetic field, as well as the geometric frustration present in structures like triangular lattices, dramatically change the properties of the system when the flux per particle is close to one flux quantum \cite{Hasegava_etal_89, Barford_Kim_91}.
However, in the case of a 2D electron system in a transverse magnetic field, the Zeeman term appears to be much larger than the chiral term originating from the orbital degrees of freedom, and therefore, magnetic polarization effects overshadow the chiral effect. Thus, the multi-spin chirality effects can be seen most prominently in the exotic case where, despite the flux acting on the orbital motion of particles, the ground state remains a spin singlet with zero net magnetization. Fortunately, ultracold atoms trapped by optical lattices represent an excellent opportunity to realize the corresponding possibilities. It has been demonstrated \cite{Delibard_etal_RMP_2011, Ketterle_etal_2013, Celi_etal_PRL_112_2014,Galitski_PT_2019} that in systems of neutral atoms on 2D optical lattices, Raman-assisted tunneling generates synthetic gauge fields and thus makes high effective fluxes penetrating the lattice without polarization of the fermion system feasible. Artificially engineered optical lattices can be generated in various geometries, including 2D triangular \cite{UCA_triangle_2010,Triangular_lattice_2019}, Kagome \cite{UCA_Kagome_Lattice_2012}, hexagonal \cite{Tarruel_12,Esslinger_13} structures as well as quasi-one-dimensional few chain systems with zig-zag \cite{Zig-Zag_ladder_2016} or standard ladder \cite{Kang_etal_18, Kang_etal_20} geometry. Moreover, lattice connectivity in optical multileg flux ladders can be ensured by extra synthetic dimensions, which can be engineered taking advantage of the internal atomic degrees of freedom \cite{Celi_etal_PRL_112_2014, Galitski_PT_2019}. 
Another advantage of the cold atoms trapped in optical lattices is the possibility to use two atom species with different masses \cite{Taglieber_etal_08}, which has opened wide possibilities in experimental realization of low-dimensional correlated fermion models with spin-dependent hopping.

Achieved progress in the simulation of artificial gauge fields opens a wide new field of condensed matter physics -- studies of the topological properties, chiral boundary currents and topological Lifshitz transitions in fermionic ladder systems with zero net magnetization in the presence of penetrating external (synthetic) magnetic flux of arbitrary intensity.
Theoretical studies of flux-ladders have been pioneered by Orignac and Giamarchi, who studied edge currents and vortex lattices in the case of interacting bosons on a standard ladder with penetrating flux \cite{Orignac_Giamarchi_01}. Later, this edge currents have been observed experimentally in the system of ultracold atoms in optical lattice \cite{Bloch_etal_2014}. Subsequent studies of the current-carrying edge states in few-chain ladder systems appear to be a good bridge for exploring correlated topological insulators with ultracold atoms \cite{Mazza_etal_2016, Zoller_etal-2017, Dalmonte_etal_2017, Dalmonte_etal_2018, Lewenstein_etal_17, Lewenstein_etal_19, Lewenstein_etal_22}. Depending on the flux and model parameters bosonic systems show insulator-superfluid and insulator-chiral superfluid transitions. 
A spin-dependent (synthetic) flux has been realised with ultracold atoms in Ref.~\cite{spin_flux}. An effective spin-dependent flux arises also in real systems with spatially varying external magnetic field or internal magnetic texture, as well as in systems with spin-orbit coupling. See Refs.~\cite{spin_gauge_1, spin_gauge_2, spin_book} for details.

The other important segment in studies of the flux ladders is connected with fermion systems. In their seminal papers Carr, Narozhny and Nersesyan \cite{NCN_05, CNN_06} show that the half-filled spinless fermion system on a standard ladder is driven to criticality by change of flux, exhibiting a very rich ground state phase diagram, including bond-ordered insulating and local current carrying orbital antiferromagnetic phases, with Berezinskii--Kosterlitz--Thouless transitions \cite{BKT_transition} between ordered and disordered ground states, and $U(1)$ Gaussian transitions between ordered phases. 
The manifold of flux driven transitions correspond to a topological Lifshitz transitions in the ground state of the system and is connected to the change in the fermionic band structure, accompanied by the change of number of Fermi points \cite{NCN_05,CNN_06, Cazalilla_etal_2022, Dalmonte_etal_2023}.

The interplay between the applied magnetic field, dynamical correlations among particles, and geometric frustration has become the subject of intense research in recent years. Main attention is focused on Fermi and Bose gases on zig-zag ladders, which are minimal quasi-one-dimensional structures that can accommodate finite magnetic fluxes through elementary plaquettes and incorporate the possibility to manipulate with both kinematic and dynamic frustration effects \cite{BN_1_EPJB, BN_2_PRB, Giamarchi_23}. These models were studied in the weak-coupling limit using the continuum-limit bosonization approach supplemented with detailed numerical calculations. It was shown that already a noninteracting system of spinless fermions on a triangular ladder exhibits cascade of flux-induced Lifshitz transitions that separate phases with the number of Fermi points equal to $4, 2, 1$ or $0$, the latter case corresponding to the band insulator phase \cite{BN_1_EPJB}. At the critical points the orbital current and its derivative with respect to flux, the charge stiffness, display singular behavior: universal square-root singularities of the ``commensurate-incommensurate" transition type when the chemical potential is constant, or cusps when a system is considered at a fixed particle density \cite{BN_1_EPJB}.
The effect of interaction has been considered in the most interesting case of half-flux quantum per triangular plaquette where the role of geometric frustration is enhanced \cite{BN_2_PRB}. Using bosonization, it was shown that in the low-energy
limit, the system is described by the quantum double--frequency sine--Gordon model. On the basis
of this correspondence, a rich phase diagram of the system is obtained. It includes trivial and
topological band insulators for weak interactions, separated by a Gaussian critical line, whereas
at larger interactions a strongly correlated phase with spontaneously broken $\mathbb{Z}_2$ symmetry sets in,
exhibiting a net charge imbalance and non-zero total current. At the intersection of the three phases,
the system features a critical point with an emergent $SU(2)$ symmetry. This non-Abelian symmetry,
absent in the microscopic description, is realized at low-energies as a combined effect of the magnetic flux, frustration, and many-body correlations.

In this paper we address the problem of flux effects in the special case of half-filled band and strong coupling. In this case, the kinematic frustration effects, well controlled by the flux in the limit of weak coupling, are partly diminished by the dominating strong interaction. We consider the most general case  of two species fermions with different masses, loaded on a lattice with a triangular ladder configuration and subjected to the \emph{spin-dependent flux} penetrating each triangular plaquette as shown in Fig.~\ref{fig:lattice}. Evolution of this system is described by the Hubbard Hamiltonian for species of atoms denoted by the "spin" index ${\sigma=\pm}$

\begin{align}
\label{eq:H_AB}
	H
	&=
	-
	\sum_{\ell \sigma}
	\Big(
		t_{A}^{\sigma}
		a^\dag_{\ell \sigma \vphantom{1}}
		a^{\vphantom{\dag}}_{\ell+1, \sigma}
		+
		\mathrm{H.c.}
	\Big)
	-
	\sum_{\ell \sigma}
	\Big(
		t_{B}^{\sigma}
		b^\dag_{\ell \sigma \vphantom{1}}
		b^{\vphantom{\dag}}_{\ell+1, \sigma}
		+
		\mathrm{H.c.}
	\Big)
\nonumber
\\
	&\mathrel{\hphantom{=}}
	-
	\sum_{\ell \sigma}
	\Big(
		t_{1}^{\sigma}
		b^\dag_{\ell \sigma \vphantom{1}}
		a^{\vphantom{\dag}}_{\ell \sigma \vphantom{1}}
		+
		\mathrm{H.c.}
	\Big)
	-
	\sum_{\ell \sigma}
	\Big(
		t_{2}^{\sigma}
		a^\dag_{\ell+1, \sigma}
		b^{\vphantom{\dag}}_{\ell \sigma \vphantom{1}}
		+
		\mathrm{H.c.}
	\Big)
	+
    U
	\sum_{\ell}
	\Big(
		n^{A}_{\ell +}
		n^{A}_{\ell -}
		+
		n^{B}_{\ell +}
		n^{B}_{\ell -}
	\Big)
\,.
\end{align}
Here $a^\dag_{\ell\sigma}$ $(a^{\vphantom{\dag}}_{\ell\sigma})$ creates (annihilates) fermion with spin \({\sigma = \pm}\) on site $\ell$ of chain $A$ (see Fig.~\ref{fig:lattice}) and \({n^{A}_{\ell\sigma} = a^\dag_{\ell\sigma} a^{\vphantom{\dag}}_{\ell\sigma}}\) is the density operator. Similarly for chain $B$.
Number of sites on both chains are equal.
We assume the periodic boundary conditions, although the derivation of the results still applies to other cases.
\begin{figure}[t!]
\centering
\begin{tikzpicture}
\newcommand{\nSegments}{2}
\newcommand{\pointDistance}{0.2*\columnwidth}
\newcommand{\xOffset}{\pointDistance/6}
\newcommand{\xmin}{-\xOffset}
\newcommand{\xmax}{\nSegments*\pointDistance + 0.5*\pointDistance + \xOffset}
\newcommand{\y}[1]{
	\ifnum#1=0 0\fi%
	\ifnum#1=1 2/3*\pointDistance\fi%
}
\newcommand{\point}[2]{({(#1 + 1/2 - #2/2) * \pointDistance}, \y{#2})}
\newcommand{\pointColor}[1]{%
	\ifnum#1=0 blue\fi%
	\ifnum#1=1 blue\fi%
}
\newcommand{\pointSize}{5pt}
\newcommand{\fluxRadius}{0.2*\y{1}}
\newcommand{\indices}[1]{
	\ifnum#1=1 $\ell-1$\fi%
	\ifnum#1=2 $\ell\vphantom{n+-1}$\fi%
	\ifnum#1=3 $\ell+1$\fi%
}
\foreach \i in {0,1}
	\draw (\xmin, \y{\i}) -- (\xmax, \y{\i});
\foreach \i in {0,...,\nSegments}
	\draw[dashed] \point{\i}{1} -- \point{\i}{0};
\foreach \i in {0,...,\number\numexpr \nSegments-1 \relax}
	\draw[dotted] \point{\i}{0} -- \point{\i+1}{1};
\foreach \j in {0,1}
	\foreach \i in {0,...,\nSegments}
		\shade[ball color=\pointColor{\j}] \point{\i}{\j} circle (\pointSize);
\foreach \i in {1,2,3} {
	\node[above=0.5em] at \point{\i-1}{1} {\indices{\i}};
	\node[below=0.5em] at \point{\i-1}{0} {\indices{\i}};
}
\node[left=0.5em] at (\xmin, \y{1}) {$A$};
\node[left=0.5em] at (\xmin, \y{0}) {$B$};
\node[above=0.25em] at (0.5*\pointDistance, \y{1}) {$t_{A}^{\sigma}$};
\node[below=0.25em] at (1*\pointDistance, \y{0}) {$t_{B}^{\sigma}$};
\node[right=0.5em] at (1.1*\pointDistance, 0.75*\y{1}) {$t_{1}^{\sigma}$};
\node[right=0.5em] at (0.6*\pointDistance, 0.25*\y{1}) {$t_{2}^{\sigma}$};
\draw[OliveGreen, thick, ->] (0.5*\pointDistance - \fluxRadius, 0.6*\y{1}) arc[start angle=-180, end angle=135, radius=\fluxRadius];
\node at (0.5*\pointDistance, 0.6*\y{1}) {$f_{A}^{\sigma}$};
\draw[OliveGreen, thick, ->] (\pointDistance - \fluxRadius, 0.4*\y{1}) arc[start angle=-180, end angle=135, radius=\fluxRadius];
\node at (\pointDistance, 0.4*\y{1}) {$f_{B}^{\sigma}$};
\end{tikzpicture}
\caption{Sketch of the lattice, where $A$ and $B$ chains are enumerated separately. There are two non-equivalent ($A$- and $B$-) triangles, carrying $f_{A}^{\sigma}$ and $f_{B}^{\sigma}$ fluxes.}
\label{fig:lattice}
\end{figure}
\noindent
The hopping parameters are complex valued and spin-dependent
\begin{align}
\label{eq:t_x_sigma}
	t_{x}^{\sigma}
	=
	\left| t_{x}^{\sigma} \right|
	\exp \! \left(\mathrm{i} \varphi_{x}^{\sigma}\right)
,
\quad
	x \in \{A, B, 1, 2\}
\,.
\end{align}
Spin-dependent phases result in a spin-dependent flux through the system. 

We restrict our consideration to the half-filled band and strong on-site interaction ${(|U| \gg |t_{x}^{\sigma}|)}$. In this case, the charge and spin degrees of freedom, for ${U>0}$ and ${U<0}$, respectively, are frozen out in the low-energy limit. Accordingly, the infrared properties of the model are described by the spin and charge degrees of freedom, respectively.
We derive the effective spin and pseudospin (charge) Hamiltonians in these limits. The whole family of  Hamiltonians is obtained, with parameters controlled by the spin-dependent hopping and flux, and containing besides the standard two-spin exchange, the Dzyaloshinskii-Moriya interaction (DMI) and complex three-spin (correlated spin-exchange) terms in the presence of an extended magnetic field.
Although below we consider the flux as a given external parameter of the model, the very presence of the chiral three-spin terms implies existence of the corresponding response in the spin system with three-spin order parameter and therefore, the possibility for the appearance of real orbital electric currents around the triangular plaquettes in the ground state of the Mott insulator \cite{Bulaevskii_etal_07, Bulaevskii_etal_09}. In this paper we do not address the question considering the ground state properties of the obtained Hamiltonians, including the above mentioned coupled spin-charge order and postpone the corresponding analysis to further studies.

The paper is organized as follows. In the forthcoming section, details of the perturbative approach, 
used to derive the effective Hamiltonian in the limit of strong on-site repulsion ${(U\gg |t^\sigma_x|)}$ are discussed.
In Sec.~\ref{sec:HubbardOperators} we briefly discuss the Hubbard operators, which are used in the subsequent Sec.~\ref{sec:Effective_Ham_U>0} to derive the effective spin Hamiltonian for ${U \gg |t^\sigma_x}|$. In Sec.~\ref{sec:Effective_Ham_U<0} the one-spin-component particle-hole transformation is used to map the model from 
${U>0}$ to ${U<0}$ case to derive the effective pseudospin (charge) Hamiltonian in the limit of strong on-site attraction ${U<0}$, ${|U|\gg |t^\sigma_x}|$. Finally, in Sec.~\ref{sec:Summary} we summarize the main results of the paper. Appendices contain technical 
details.
Technical details of the derivation of the effective Hamiltonian are summarized in App.~\ref{sec:App_H_eff_derivation}, while the details of the gauge transformations are discussed in App.~\ref{sec:App_gauge}.

\section{The model}

In what follows, we consider the half-filled band case and strong on-site interaction ${(U \gg |t_{x}^{\sigma}|)}$. In this limit the perturbative treatment
of the half-filled Hubbard model based on expansion of the Hamiltonian in powers of $|t_{x}^{\sigma}|/U$ goes back to the original derivation of the effective spin Hamiltonian to the second order by Anderson~\cite{Anderson59}.
Later expansion up to the fourth-order terms has been obtained by Bulaevskii~\cite{Bulaevskii67} 
and Takahashi~\cite{Takahashi77}. 

An alternative approach to construct the effective Hamiltonian is based on unitary transformations. Harris and Lange~\cite{Harris_Lange_67} used such a transformation to obtain second-order results and to calculate spectral properties of the Hubbard model.  A consistent transformation scheme which allows one to remove all unphysical terms and to derive the $t/U$-expansion up to any desired order has been formulated by MacDonald, Girvin and Yoshioka~\cite{MacDonald88}.
In their scheme, interaction terms which do not conserve the number of local electron pairs are eliminated from the Hamiltonian order by order in an iterative treatment, generating new interactions and thus improving the accuracy of the transformation at each step. Below we apply the method developed in Ref.~\cite{MacDonald88} (see also~\cite{GMJ_2016})
to the Hubbard model on a zig-zag ladder with \emph{spin-asymmetric hopping and spin-dependent penetrating flux} given by the Hamiltonian~\eqref{eq:H_AB}. 

For performing calculations, it is convenient to enumerate the entire system along the inter-chain zig-zag line (see Fig.~\ref{fig:lattice_zigzag}) and rewrite the Hamiltonian in an alternative compact form
\begin{align}
\label{eq:H_ij}
	H
	=
	T + V
	=
	-\sum_{ij\sigma}
		t_{ij}^{\sigma}
		c^\dag_{i\sigma}
		c^{\vphantom{\dag}}_{j\sigma}
	+
	U
    \sum_{i}
        n_{i+}
		n_{i-}
\,.
\end{align}
Here \({c_{2\ell-1,\sigma} = a_{\ell\sigma}}\) and \({c_{2\ell,\sigma} = b_{\ell\sigma}}\). 
Hopping parameters in Eqs.~\eqref{eq:H_AB} and \eqref{eq:H_ij} relate as 
\begin{align}
\label{eq:t_ij_to_x}
	t_{ij}^{\sigma}
	=
	\begin{cases}
		t_{1}^{\sigma} &\text{for $i = 2\ell$, $j = 2\ell - 1$} \\
		t_{2}^{\sigma} &\text{for $i = 2\ell + 1$, $j = 2\ell$} \\
		t_{A}^{\sigma} &\text{for $i = 2\ell - 1$, $j = 2\ell + 1$} \\		
		t_{B}^{\sigma} &\text{for $i = 2\ell$, $j = 2\ell + 2$}
	\end{cases}
\end{align}
and \({t_{ij}^\sigma = 0}\) for \({|i - j| > 2}\).
Remaining non-zero elements are obtained by complex conjugating Eq.~\eqref{eq:t_ij_to_x} and using the Hermitian property of the hopping matrix for each $\sigma$, \({\left(t_{ij}^{\sigma}\right)^{\! *} = t_{ji}^{\sigma}}\). Hence,
\begin{align}
	\left|t_{ji}^{\sigma}\right| = \left|t_{ij}^{\sigma}\right|
,\quad
	\varphi_{ji}^{\sigma} = -\varphi_{ij}^{\sigma}
\,.
\end{align}

In the considered case of half-filled band, number of particles with spin $\sigma$ is ${N_{\sigma}=N/2}$, where $N$ is the number of lattice sites.
There are two non-equivalent triangular plaquettes, which we will refer to as $A$- and $B$-triangles. They carry the fluxes
\begin{subequations}
\label{eq:fluxes}
\begin{align}
	f_{A}^{\sigma} &= \varphi_{A}^{\sigma} + \varphi_{1}^{\sigma} + \varphi_{2}^{\sigma}
\,,\\
	f_{B}^{\sigma} &= -\left(\varphi_{B}^{\sigma} + \varphi_{1}^{\sigma} + \varphi_{2}^{\sigma}\right)
,
\end{align}
\end{subequations}
in the units of flux quantum $hc/e$, which we set to 1. 
We parametrize the absolute values of the hopping parameters as
\begin{align}
	\left| t_{ij}^{+} \right|
	\equiv
	\tau_{ij}
	\lambda
\,,
\quad
	\left| t_{ij}^{-} \right|
	\equiv
	\tau_{ij}
	/
	\lambda
\,.
\end{align}
Here ${\tau_{ij}=\left| t_{ij}^{+} t_{ij}^{-} \right|^{1/2}}$ is the arithmetic-geometric mean, while ${\lambda = \left| t_{ij}^{+} / t_{ij}^{-} \right|^{1/2}}$ (restricted to be equal for all links $ij$) is a measure of kinetic spin-asymmetry. The value \({\lambda = 1}\) indicates spin-independence, \({\left| t_{ij}^{+} \right| = \left| t_{ij}^{-} \right|}\). By construction, $\tau_{ij}$ and $\lambda$ are real and positive. 

Without loss of generality, one can assume that \({\varphi_1^\pm = \varphi_2^\pm = 0}\), \({\varphi_A^\pm = f_A^\pm}\) and \({\varphi_B^\pm = -f_B^\pm}\). 
Although, there exists a more convenient gauge if the condition \({f_A^+ + f_B^+ = \pm(f_A^- + f_B^-)}\) is satisfied. 
See App.~\ref{sec:App_gauge} for more details on gauge transformation.
For this cause, we consider arbitrary \({\varphi_x^\sigma}\).

The kinetic part can be decomposed into three parts, \({T = T_{0} + T_{1} + T_{-1}}\), where $T_{m}$ changes the number of doubly occupied sites by $m$
\begin{subequations}
\label{eq:T_m_ij}
\begin{align}
	T_{0}
	&=
	-\sum_{i j \sigma}
		t_{ij}^{\sigma}
		\mathopen{}\left(
			n^{\vphantom{\dag}}_{i \bar{\sigma}}
			c^\dag_{i \sigma}
			c^{\vphantom{\dag}}_{j \sigma}
			n^{\vphantom{\dag}}_{j \bar{\sigma}}
			+
			h^{\vphantom{\dag}}_{i \bar{\sigma}}
			c^\dag_{i \sigma}
			c^{\vphantom{\dag}}_{j \sigma}
			h^{\vphantom{\dag}}_{j \bar{\sigma}}
		\right)
,
\\
	T_{1}
	&=
	-\sum_{i j \sigma}
		t_{ij}^{\sigma}
		n^{\vphantom{\dag}}_{i \bar{\sigma}}
		c^\dag_{i \sigma}
		c^{\vphantom{\dag}}_{j \sigma}
		h^{\vphantom{\dag}}_{j \bar{\sigma}}
\,,
\\
	T_{-1}
	&=
	-\sum_{i j \sigma}
		t_{ij}^{\sigma}
		h^{\vphantom{\dag}}_{i \bar{\sigma}}
		c^\dag_{i \sigma}
		c^{\vphantom{\dag}}_{j \sigma}
		n^{\vphantom{\dag}}_{j \bar{\sigma}}
\,,
\end{align}
\end{subequations}
where \({h_{i \sigma} = 1 - n_{i \sigma}}\) and ${\bar{\sigma}=-\sigma}$. They satisfy \({T^\dag_{m} = T^{\vphantom{\dag}}_{-m}}\) and most importantly
\begin{align}
\label{eq:commutator_VTm}
	\left[V , T_{m}\right]
	=
	m U T_{m}
\,.
\end{align}

\begin{figure}[t!]
\centering
\begin{tikzpicture}
\newcommand{\nSegments}{2}
\newcommand{\pointDistance}{0.2*\columnwidth}
\newcommand{\xOffset}{\pointDistance/6}
\newcommand{\xmin}{-\xOffset}
\newcommand{\xmax}{\nSegments*\pointDistance + 0.5*\pointDistance + \xOffset}
\newcommand{\y}[1]{
	\ifnum#1=0 0\fi%
	\ifnum#1=1 2/3*\pointDistance\fi%
}
\newcommand{\point}[2]{({(#1 + 1/2 - #2/2) * \pointDistance}, \y{#2})}
\newcommand{\pointColor}[1]{%
	\ifnum#1=0 blue\fi%
	\ifnum#1=1 blue\fi%
}
\newcommand{\pointSize}{5pt}
\newcommand{\fluxRadius}{0.2*\y{1}}
\newcommand{\indicesA}[1]{
	\ifnum#1=1 $2\ell-3$\fi%
	\ifnum#1=2 $2\ell-1\vphantom{n+-1}$\fi%
	\ifnum#1=3 $2\ell+1$\fi%
}
\newcommand{\indicesB}[1]{
	\ifnum#1=1 $2\ell-2$\fi%
	\ifnum#1=2 $2\ell\vphantom{n+-1}$\fi%
	\ifnum#1=3 $2\ell+2$\fi%
}
\foreach \i in {0,1}
	\draw (\xmin, \y{\i}) -- (\xmax, \y{\i});
\foreach \i in {0,...,\nSegments}
	\draw[dashed] \point{\i}{1} -- \point{\i}{0};
\foreach \i in {0,...,\number\numexpr \nSegments-1 \relax}
	\draw[dotted] \point{\i}{0} -- \point{\i+1}{1};
\foreach \j in {0,1}
	\foreach \i in {0,...,\nSegments}
		\shade[ball color=\pointColor{\j}] \point{\i}{\j} circle (\pointSize);
\foreach \i in {1,2,3} {
	\node[above=0.5em] at \point{\i-1}{1} {\indicesA{\i}};
	\node[below=0.5em] at \point{\i-1}{0} {\indicesB{\i}};
}
\node[left=0.5em] at (\xmin, \y{1}) {$A$};
\node[left=0.5em] at (\xmin, \y{0}) {$B$};
\node[above=0.25em] at (0.5*\pointDistance, \y{1}) {$t_{A}^{\sigma}$};
\node[below=0.25em] at (1*\pointDistance, \y{0}) {$t_{B}^{\sigma}$};
\node[right=0.5em] at (1.1*\pointDistance, 0.75*\y{1}) {$t_{1}^{\sigma}$};
\node[right=0.5em] at (0.6*\pointDistance, 0.25*\y{1}) {$t_{2}^{\sigma}$};
\draw[OliveGreen, thick, ->] (0.5*\pointDistance - \fluxRadius, 0.6*\y{1}) arc[start angle=-180, end angle=135, radius=\fluxRadius];
\node at (0.5*\pointDistance, 0.6*\y{1}) {$f_{A}^{\sigma}$};
\draw[OliveGreen, thick, ->] (\pointDistance - \fluxRadius, 0.4*\y{1}) arc[start angle=-180, end angle=135, radius=\fluxRadius];
\node at (\pointDistance, 0.4*\y{1}) {$f_{B}^{\sigma}$};
\end{tikzpicture}
\caption{Sketch of the lattice, where the entire system is enumerated along the inter-chain zig-zag line. Site indices on chains $A$ and $B$ are odd and even, respectively.}
\label{fig:lattice_zigzag}
\end{figure}

\section{\texorpdfstring{The large $U$ limit}{The large U limit}}
Our aim is to derive the low-energy effective Hamiltonian from the exact Hamiltonian~\eqref{eq:H_ij} using the Schrieffer-Wolff (SW) method, where the low-energy effective Hamiltonian is obtained from the exact Hamiltonian by a unitary transformation decoupling the low- and high-energy subspaces~\cite{SW_66}. We follow the scheme  based on SW transformation and formulated in Ref.~\cite{MacDonald88}, with slightly modified notation. 

\subsection{Schrieffer-Wolff transformation}

In the large-$U$ limit of the standard Hubbard model the many-electron states are grouped according to the number of doubly occupied sites (doublons) $N_{d}$. The hopping operator $T$ mixes the states with different $N_{d}$.  The "unmixing" can be achieved by introducing suitable linear combinations of the uncorrelated basis states.  
The $\mathcal{S}$ matrix for this transformation, and the transformed Hamiltonian,
\begin{align}
	H_{\mathrm{eff}} = e^{\mathcal{S}} H e^{-\mathcal{S}} \,,
\end{align}
are generated by an iterative procedure, which results in an
expansion in powers of the hopping amplitudes $t_{ij}^{\sigma}$ divided by the on-site interaction energy $U$
\begin{align}
\label{eq:H'}
	H_{\mathrm{eff}}
	=
	e^{\mathcal{S}} H e^{-\mathcal{S}}
	=
	H
	+
	\left[\mathcal{S}, H\right]
	+
	\frac{1}{2!}
	\left[\mathcal{S}, \left[\mathcal{S}, H\right]\right]
	+
	\cdots
\end{align}
We seek $\mathcal{S}$ such that $H_{\mathrm{eff}}$ contains only the terms that preserve the number of doubly occupied sites, up to the desired order in $t_{ij}^\sigma/U$.
Denoting the expansion
\begin{align}
	\mathcal{S}
	=
	\sum_{k=1}^{\infty}
		\frac{\mathcal{S}_{k}}{U^{k}}
\,,
\end{align}
the higher order coefficient $\mathcal{S}_{k}$ can be recursively determined from the lower order ones $\mathcal{S}_{1}$, $\mathcal{S}_{2}$, $\cdots$, $\mathcal{S}_{k-1}$. The relation that underpins the existence of such a scheme is derived from Eq.~\eqref{eq:commutator_VTm}
\begin{align}
	\big[V, T^{(k)}_{\bm{m}}\big]
	=
	U
	M^{(k)}_{\bm{m}}
	T^{(k)}_{\bm{m}}
\,,
\qquad
	M^{(k)}_{\bm{m}}
	=
	\sum_{i=1}^{k}
		m_{i}
\,,
\end{align}
where $\bm{m}$ denotes the vector \({(m_1, m_2, \ldots, m_k)}\) with \({m_{i} = 0, \pm 1}\) and $T^{(k)}_{\bm{m}}$ is the product of several $T_{m}$ operators
\begin{align}
\label{eq:T^k_m}
	T^{(k)}_{\bm{m}}
	:=
	T_{m_{1}}
	T_{m_{2}}
	\cdots
	T_{m_{k}}
\,.
\end{align}
Any term that changes the number of doubly occupied sites, \({U^{-(k-1)} T^{(k)}_{\bm{m}}}\) with \({M^{(k)}_{\bm{m}} \neq 0}\), can be eliminated by adding \({T^{(k)}_{\bm{m}} / M^{(k)}_{\bm{m}}}\) to $\mathcal{S}_{k}$. In the end, the commutator \({\left[\mathcal{S}_{k}, V\right]}\) will cancel all such terms, while the remaining commutators in $H_{\mathrm{eff}}$, Eq.~\eqref{eq:H'}, will produce higher order terms.
Conforming to this procedure, we obtain
\begin{subequations}
\begin{align}
	\mathcal{S}_{1}
	&=
	T_{1} - T_{-1}
\,,
\\
	\mathcal{S}_{2}
	&=
	\left[T_{1} , T_{0}\right]
	+
	\left[T_{-1} , T_{0}\right]
,
\\
	\mathcal{S}_{3}
	&=
	\left[T_{0}, \left[T_{0}, T_{1}\right]\right]
	-
	\left[T_{0}, \left[T_{0}, T_{-1}\right]\right]
    +
	\frac{2}{3}
	\Big(
		\left[T_{1}, \left[T_{1}, T_{-1}\right]\right]
		-
		\left[T_{-1}, \left[T_{-1}, T_{1}\right]\right]
	\Big)
\nonumber
\\
    &\hphantom{=}
    +
	\frac{1}{4}
	\Big(
		\left[T_{-1}, \left[T_{-1}, T_{0}\right]\right]
		-
		\left[T_{1}, \left[T_{1}, T_{0}\right]\right]
	\Big)
.
\end{align}
\end{subequations}
This allows us to ensure the desired property up to $H_{\mathrm{eff}}^{(3)}$ in the expansion
\begin{align}
	H_{\mathrm{eff}}
	=
	V
	+
	\sum_{k=1}^{\infty}
		H_{\mathrm{eff}}^{(k)}
\,,
\qquad
	H_{\mathrm{eff}}^{(k)}
	\equiv
	\frac{\overline{H}_{\mathrm{eff}}^{(k)}}{U^{k-1}}
\,,
\end{align}
which corresponds to $3^{\mathrm{rd}}$ order in $t/U$, relative to $V$. 
The expansion coefficients $\overline{H}_{\mathrm{eff}}^{(k)}$ do not depend on $U$
\begin{subequations}
\begin{align}
	\overline{H}_{\mathrm{eff}}^{(1)}
	&=
	T_{0}
\,,
\\
	\overline{H}_{\mathrm{eff}}^{(2)}
	&=
	\left[T_{1} , T_{-1}\right]
,
\\
	\overline{H}_{\mathrm{eff}}^{(3)}
	&=
	\frac{1}{2}
	\Big(
		\left[T_{1}, \left[T_{0}, T_{-1}\right]\right]
		+
		\left[T_{-1}, \left[T_{0}, T_{1}\right]\right]
	\Big)
.
\end{align}
\end{subequations}
Transformed Hamiltonian can be further simplified on the projected Hilbert space and even further in the considered case of half-filled band. Next section is dedicated to this program.

\subsection{The Half-filled band case and strong repulsive interaction}
\label{sec:H_eff_T}

In this section we focus on the case of a half-filled band, where in the large-$U$ limit the lowest energy state $\left| \psi_{L} \right>$ has exactly one electron at each site. In this subspace no hops are possible without increasing the number of doubly occupied sites. Therefore, we have
\begin{align}
\label{eq:Tm_GS_U>0}
	V \left| \psi_{L} \right> = 0 \,,\qquad
	T_{0} \left| \psi_{L} \right> = 0 \,,\qquad
	T_{-1} \left| \psi_{L} \right> = 0 \,.
\end{align}
Note that the last relation is valid for any filling, while the first two are exclusive to half-filling, \({N_{+} + N_{-} = N}\), where $N$ is the number of lattice sites. 
In fact, there are more terms that produce zero when acting on \({\left| \psi_{L} \right>}\). Consider an arbitrary splitting of some product operator~\eqref{eq:T^k_m} into two parts
\begin{align}
\label{eq:split_Tkm}
	T^{(k)}_{\bm{m}}
	=
	T^{(k')}_{\bm{m'}}
	T^{(k'')}_{\bm{m''}}
\,,
\qquad
	\left(k'+k''=k\right)
.
\end{align}
If there exists the splitting \({1 \leqslant k' \leqslant k}\), such that \({M^{(k'')}_{\bm{m''}} = 0}\) and \({m'_{k'} \neq 1}\), then \({T^{(k)}_{\bm{m}} \left| \psi_L \right> = 0}\). Discarding such terms, the transformed Hamiltonian boils down to
\begin{subequations}
\label{eq:H'_U>0}
\begin{align}
	\overline{\mathcal{H}}_{\mathrm{eff}}^{(1)}
	&=
	0
\,,
\\
\label{eq:H'_2}
	\overline{\mathcal{H}}_{\mathrm{eff}}^{(2)}
	&=
	-T_{-1} T_{1}
\,,
\\
\label{eq:H'_3}
	\overline{\mathcal{H}}_{\mathrm{eff}}^{(3)}
	&=
	T_{-1} T_{0} T_{1}
\,.
\end{align}
\end{subequations}
Here and onward, calligraphic $\mathcal{H}$ indicates that the effective Hamiltonian is restricted to the ground state of the half-filled band.

\section{Hubbard operators}
\label{sec:HubbardOperators}

To handle the effects of strong repulsive (attractive) interaction properly, it is
important to know whether at the beginning or at the end of a given
hopping process a particular site is doubly (singly) occupied or not. For
this purpose one introduces the so-called Hubbard operators~\cite{Hubbard-Operators-Book}, 
which are defined at each site of the lattice and describe all possible transitions
between the local basis states 
\begin{align}
	X^{ab}_{i}
	=
	| a \rangle
	\langle b |
\,,
\end{align}
where \({a,b \in \left\{0, +, -, 2\right\}}\), and the site index $i$ is not indicated in the right hand side for brevity. To clarify the fermionic sign convention, under \({| 2 \rangle}\) we mean \({| 2 \rangle \equiv c^\dag_{i +} c^\dag_{i -} | 0 \rangle}\).

The original fermion creation (annihilation) operators can be expressed in terms of Hubbard operators as
\begin{align}
	c^\dag_{i \sigma}
	=
	X^{\sigma 0}_{i}
	+
	\sigma
	X^{2 \bar{\sigma}}_{i}
\,,
\quad
	c^{\vphantom{\dag}}_{i \sigma}
	=
	X^{0 \sigma}_{i}
	+
	\sigma
	X^{\bar{\sigma} 2}_{i}
\,.
\end{align}
Conversely, $X$-operators can be expressed as
\begin{subequations}
\label{eq:X_to_c}
\begin{alignat}{4}
	&X^{2 \sigma}_{i}
	&&=
	\bar{\sigma}
	n^{\vphantom{\dag}}_{i \sigma}
	c^\dag_{i \bar{\sigma}}
\,,
\qquad
	&&X^{\sigma 2}_{i}
	&&=
	\bar{\sigma}
	c^{\vphantom{\dag}}_{i \bar{\sigma}}
	n^{\vphantom{\dag}}_{i \sigma}
\,,
\\
	&X^{\sigma 0}_{i}
	&&=
	h^{\vphantom{\dag}}_{i \bar{\sigma}}
	c^\dag_{i \sigma}
\,,
\qquad
	&&X^{0 \sigma}_{i}
	&&=
	c^{\vphantom{\dag}}_{i \sigma}
	h^{\vphantom{\dag}}_{i \bar{\sigma}}
\,,
\\
	&X^{\sigma \sigma}_{i}
	&&=
	n_{i \sigma}
	h_{i \bar{\sigma}}
\,,
\qquad
	&&X^{\sigma \bar{\sigma}}_{i}
	&&=
	c^\dag_{i \sigma}
	c^{\vphantom{\dag}}_{i \bar{\sigma}}
\,,
\\
	&X^{2 0}_{i}
	&&=
	c^\dag_{i +}
	c^\dag_{i -}
\,,
\qquad
	&&X^{0 2}_{i}
	&&=
	c^{\vphantom{\dag}}_{i -}
	c^{\vphantom{\dag}}_{i +}
\,,
\\
	&X^{0 0}_{i}
	&&=
	h_{i +}
	h_{i -}
\,,
\qquad
	&&X^{2 2}_{i}
	&&=
	n_{i +}
	n_{i -}
\,.
\end{alignat}
\end{subequations}

Several properties of $X$-operators are manifest
\begin{subequations}
\begin{gather}
	\sum_{a} X^{aa}_{i} = 1
\,,
\qquad
	\left(X^{ab}_{i}\right)^\dag = X^{ba}_{i}
\,,
\\
\label{eq:X_contraction}
	X^{a_{1} b_{1}}_{i}
	X^{a_{2} b_{2}}_{i}
	=
	\delta_{a_{2} b_{1}}
	X^{a_{1} b_{2}}_{i}
\,.
\end{gather}
\end{subequations}

Out of these 16 operators, 8 $(X^{2 \sigma}_{i}$, $X^{\sigma 2}_{i}$, $X^{\sigma 0}_{i}$, $X^{0 \sigma}_{i})$ are of the Fermi type and the remaining 8 $(X^{\sigma \sigma}_{i}$, $X^{\sigma \bar{\sigma}}_{i}$, $X^{2 0}_{i}$, $X^{0 2}_{i}$, $X^{0 0}_{i}$, $X^{2 2}_{i})$ are of the Bose type.
Two $X$-operators of the Fermi type satisfy the anti-commutation relations
\begin{subequations}
\label{eq:X_comm_anticomm_rules}
\begin{align}
	\left\{
		X^{a_{1} b_{1}}_{i} ,
		X^{a_{2} b_{2}}_{j}
	\right\}
	&=
	\delta_{ij}
	\left(
		\delta_{a_{2} b_{1}}
		X^{a_{1} b_{2}}_{i}
		+
		\delta_{a_{1} b_{2}}
		X^{a_{2} b_{1}}_{i}
	\right)
.
\intertext{Two $X$-operators, where at least one of them is of Bose type, satisfy the commutation relations}
	\left[
		X^{a_{1} b_{1}}_{i} ,
		X^{a_{2} b_{2}}_{j}
	\right]
	&=
	\delta_{ij}
	\left(
		\delta_{a_{2} b_{1}}
		X^{a_{1} b_{2}}_{i}
		-
		\delta_{a_{1} b_{2}}
		X^{a_{2} b_{1}}_{i}
	\right)
.
\end{align}
\end{subequations}

It is straightforward to represent the hopping terms~\eqref{eq:T_m_ij} by the
Hubbard operators using Eqs.~\eqref{eq:X_to_c}
\begin{subequations}
\label{eq:T_to_X}
\begin{align}
\label{eq:T0_to_X}
	T_{0}
	&=
	-\sum_{i j \sigma}
		t_{ij}^{\sigma}
		\left(
			X^{2 \bar{\sigma}}_{i}
			X^{\bar{\sigma} 2}_{j}
			+
			X^{\sigma 0}_{i}
			X^{0 \sigma}_{j}
		\right)
,
\\
\label{eq:T+_to_X}
	T_{1}
	&=
	-\sum_{i j \sigma}
		\sigma
		t_{ij}^{\sigma}
		X^{2 \bar{\sigma}}_{i}
		X^{0 \sigma}_{j}
,
\\
\label{eq:T-_to_X}
	T_{-1}
	&=
	-\sum_{i j \sigma}
		\sigma
		t_{ij}^{\sigma}
		X^{\sigma 0}_{i}
		X^{\bar{\sigma} 2}_{j}
.
\end{align}
\end{subequations}

One also easily verifies that the $X$-operators describing the transitions between
singly occupied states can be rewritten in terms of spin ${S = 1/2}$ operators as
\begin{subequations}
\label{eq:X-to-S}
\begin{gather}
\label{eq:X-to-S-pm}
	X_i^{+ -}
	=
	c^\dag_{i +}
	c^{\vphantom{\dag}}_{i -}
	=
	S^+_i
\,,
\qquad 
	X_i^{- +}
	=
	c^\dag_{i -}
	c^{\vphantom{\dag}}_{i +}
	=
	S^-_i
\,,
\\
	X_i^{+ +}
	=
	\frac{1}{2}
	+
	S^z_i
\,,
\qquad
	X_i^{- -}
	=
	\frac{1}{2}
	-
	S^z_i
\,.
\end{gather}
\end{subequations}
Similarly the $X$-operators describing the transitions between doubly occupied and empty states can be rewritten in terms of pseudospin (charge) $\eta$ operators, which satisfy commutation rules of spin ${S = 1/2}$ operators~\cite{Micnas_etal_90}
\begin{subequations}
\label{eq:X-to-Eta}
\begin{gather}
	X_i^{2 0}
	=
	c^\dag_{i +}
	c^\dag_{i -}
	=
	\eta^+_i
\,,
\qquad
	X_i^{0 2}
	=
	c^{\vphantom{\dag}}_{i -}
	c^{\vphantom{\dag}}_{i +}
	=
	\eta^-_i
\,,
\\
	X_i^{2 2}
	=
	n_{i +}
	n_{i -}
	=
	\frac{1}{2}
	+
	\eta^z_i
\,,
\qquad
	X_i^{0 0}
	=
	h_{i +}
	h_{i -}
	=
	\frac{1}{2}
	-
	\eta^z_i
\,,
\end{gather}
\end{subequations}
where ${\eta^z_i = (n_{i +} + n_{i -} - 1) / 2}$. 

To summarize, the relations~\eqref{eq:X-to-S} and \eqref{eq:X-to-Eta} allow us to rewrite the effective Hamiltonian $\mathcal{H}_{\mathrm{eff}}$, expressed  in terms of the Hubbard operators, in a final and the most compact form in terms of spin \({S=1/2}\) for repulsive interaction and the pseudospin (charge) \({\eta=1/2}\) operators in the case of strong attractive coupling, describing respectively the low-energy sectors of spin and charge excitations.

\section{Effective spin \texorpdfstring{$S=1/2$}{S=1/2} Hamiltonian in the case of strong repulsive interaction}
\label{sec:Effective_Ham_U>0}

Utilizing Eqs.~\eqref{eq:T_to_X} and \eqref{eq:X-to-S} in Eqs.~\eqref{eq:H'_U>0}, we obtain effective spin \({S = 1/2}\) Hamiltonian. (Technical details of the derivation are provided in App.~\ref{sec:App_H_eff_derivation})
\begin{align}
\label{eq:H_eff_Spin}
	\mathcal{H}_{\mathrm{eff}}
	&=
	\sum_{i < j}
		J_{ij}
		\Big[
			\frac{1}{2}
			\left(
				e^{\mathrm{i} \theta^{-}_{ij}}
				S^{+}_{i}
				S^{-}_{j}
				+
				\mathrm{H.c.}
			\right)
			+
			\gamma
			S^{z}_{i}
			S^{z}_{j}
		\Big]
\nonumber
\\
	&\mathrel{\hphantom{=}}
	\mathrel{+}
	\sum_{i < j < k}
		J_{ijk}^{\vphantom{-}}	
		\bigg\{
			h_{ijk}^{-}
			\left(S^{z}_i + S^{z}_j + S^{z}_k-12S^{z}_i S^{z}_j S^{z}_k\right)
\nonumber
\\
			&\mathrel{\hphantom{=}} \qquad\qquad\qquad
			\mathrel{+}
			6
			\Big[
				w_{ijk}^{-}
				\left(
					e^{\mathrm{i} \theta^{-}_{ij}}
					S^{+}_i
					S^{-}_j
					S^{z}_k
					+
					e^{\mathrm{i} \theta^{-}_{jk}}
					S^{+}_j
					S^{-}_k
					S^{z}_i 
					+
					e^{\mathrm{i} \theta^{-}_{ki}}
					S^{+}_k
					S^{-}_i
					S^{z}_j
				\right)
				+
				\mathrm{H.c.}
			\Big]
		\bigg\}
\,,
\end{align}
with
\begin{align}
\label{eq:J_ij}
	J_{ij}
	&=
	\frac{4\tau_{ij}^2}{U}
\,,
\qquad
	\gamma
	=
	\left(\lambda^{2} + \lambda^{-2}\right) \! / 2
\,,
\quad
	\left(\gamma \geqslant 1\right)
,
\\
\label{eq:J_ijk}
	J_{ijk}
	&=
	\frac{\tau_{ij} \tau_{jk} \tau_{ki}}{U^2}
\,,
\\
\label{eq:theta_ijk}
	\theta^{-}_{ij}
	&=
	\varphi_{ij}^{+} - \varphi_{ij}^{-}
\,.
\end{align}
Here \({\tau_{ij} = |t_{ij}^{+} t_{ij}^{-}|^{1/2}}\), \({\lambda = |t_{ij}^{+} / t_{ij}^{-}|^{1/2}}\) is independent of $i,j$ indices and  
\begin{align}
\label{eq:h_ijk}
	h_{ijk}^{-}
	&=
	\lambda^{3}
	\cos \! \left(f_{ijk}^{+}\right)
	-
	\lambda^{-3}
	\cos \! \left(f_{ijk}^{-}\right)
,
\\
\label{eq:w_ijk}
	w_{ijk}^{-}
	&=
	\lambda^{-1}
	\exp \! \left(-\mathrm{i} f_{ijk}^{-}\right)
	-
	\lambda
	\exp \! \left(\mathrm{i} f_{ijk}^{+}\right)
.
\end{align}
The phase $f_{ijk}^{\sigma}$ is equal to either $f_{A}^{\sigma}$ or $-f_{B}^{\sigma}$, depending on which triangle the indices $i,j,k$ (with \({i<j<k}\)) form. 

It is instructive to consider several limiting cases where $\mathcal{H}_{\mathrm{eff}}$ acquires a simpler form, in one way or another.
However, analysis of the ground state properties of the corresponding models is out of the scope of this paper.

Let us first focus on the two-spin part of the Hamiltonian \eqref{eq:H_eff_Spin} which constitutes the $XXZ$ Heisenberg model with restricted anisotropy \({\gamma \geqslant 1}\) and \emph{complex exchange interaction} in $XY$ plane. The imaginary part of the complex phase factor of the transverse exchange interaction corresponds to the DMI in the system, with orientation vector along $z$-axis,
\begin{align}
\label{eq:DMI}
	\frac{J_{ij}}{2}
	\left(\mathrm{i} \sin\theta^{-}_{ij} \right)
	S^{+}_{i}
	S^{-}_{j}
	+
	\mathrm{H.c.}
	=
	D_{ij}^{z}
	\left(\bm{S}_{i} \times \bm{S}_{j}\right)^{z}
,
\qquad
	D_{ij}^{z}
	=
	J_{ij}
	\sin\theta^{-}_{ij}
\,.
\end{align}
The complex phase factors, hence the DMI, can be eliminated by gauge transformation if and only if the fluxes are spin-independent modulo $\pi$, i.e.\
\({f_{ijk}^{+} = f_{ijk}^{-} \mod{\pi}}\) (see App.~\ref{sec:App_gauge_corollary}).
We emphasize that the DMI emerges purely due to spin-dependent flux.
In terms of original fermions, the DMI in Eq.~\eqref{eq:DMI} describes a spin-current
\begin{align}
\label{eq:spin_current}
	\left(
		\bm{S}_{i}
		\times
		\bm{S}_{j}
	\right)^{z}
	\sim
	\mathrm{i}
	\left(
		c^\dag_{i+}
		c^\dag_{j-}
		c^{\vphantom{\dag}}_{i-}
		c^{\vphantom{\dag}}_{j+}
		-
		\mathrm{H.c.}
	\right)
.
\end{align}
Here we used Eq.~\eqref{eq:X-to-S-pm} and ${S^\pm_i = S^x_i \pm \mathrm{i} S^y_i}$.
The physical explanation is as follows. Within the projected Hilbert space at half-filling, where there is exactly one particle per site, the only allowed process is the exchange of particles on different sites. During such exchanges, charge is not transferred. Spin is transferred when a single exchange of $\sigma$ and $-\sigma$ particles occurs, but they happen in opposite directions. For spin-independent flux, average rate in both directions is the same, whereas for spin-dependent flux, it is not. This imbalance is exactly what Eqs.~\eqref{eq:DMI} and \eqref{eq:spin_current} describe.

\subsection{Limiting cases}
\label{sec:limits}

We begin this section by considering under what conditions the three-spin terms are absent. 
If one of the hopping amplitudes is zero in each triangular plaquette, then ${J_{ijk} = 0}$ for all $ijk$ and hence ${\mathcal{H}^{(3)}_{\mathrm{eff}} \equiv 0}$.
In particular, if the intra-chain hopping amplitudes vanish, \({t_A^\pm = t_B^\pm = 0}\), then the system is equivalent to a simple spin chain. While if one of the inter-chain hopping amplitudes vanish, ${t_1^\pm = 0}$ or ${t_2^\pm = 0}$, then the system constitutes a square ladder.

Assuming ${J_{ijk} \neq 0}$, the condition ${w^{-}_{ijk} = 0}$ is fulfilled if and only if the hopping amplitudes are spin-independent, ${\lambda = 1}$, and the total flux per triangle vanishes, ${f^{+}_{ijk} + f^{-}_{ijk} = 0}$. It is straightforward to see that in such case the magnetic field vanishes too, ${h^{-}_{ijk} = 0}$.
Below we consider the cases where either ${\lambda \neq 1}$ or/and ${f^{+}_{ijk} + f^{-}_{ijk} \neq 0}$ and thus, the three-spin terms are present.

\subsubsection*{Spin-symmetric case}
In a spin-symmetric case, i.e.\ \({\lambda = 1}\) and \({f_{ijk}^{+} = f_{ijk}^{-} \equiv f_{ijk}}\), the phase factors \({\theta^{-}_{ij} = 0}\) for any $i$ and $j$, \({\gamma = 1}\), \({h^{-}_{ijk} = 0}\) and $w^{-}_{ijk}$ are purely imaginary. In that case $\mathcal{H}^{(2)}_{\mathrm{eff}}$ coincides with $XXX$ spin chain Hamiltonian, while $\mathcal{H}^{(3)}_{\mathrm{eff}}$ reduces to only a scalar spin chirality term
\begin{align}
\label{eq:H_eff_Spin_symmetric}
	\mathcal{H}_{\mathrm{eff}}
	=
	\sum_{i < j}
		J_{ij}
		\Big[
			\frac{1}{2}
			\left(
				S^{+}_{i}
				S^{-}_{j}
				+
				\mathrm{H.c.}
			\right)
			+
			S^{z}_{i}
			S^{z}_{j}
		\Big]
	-
	24
	\sum_{i < j < k}
		J_{ijk}
		\sin \! \left(f_{ijk}\right)
		\bm{S}_{i}
		\cdot
		\left[
			\bm{S}_{j}
			\times
			\bm{S}_{k}
		\right]
,
\end{align}
obtained earlier in Ref.~\cite{Sen-Chjtra_95}. 
Obviously, the scalar spin chirality term vanishes together with the flux, \({f_{ijk} = 0}\) mod $\pi$.
This model is exactly solvable by the Bethe ansatz at the special point in the parameter space. In Ref.~\cite{Frahm_1997}, the magnetic phase diagram was obtained in ${\kappa - h}$ plane, where the single parameter $\kappa$ determines all the coupling constants in the special case and $h$ is the magnetic field. Numerical (density matrix renormalization group) analysis of the correlation functions for the Hamiltonian, consisting of the scalar spin chirality terms only, has been done in Ref.~\cite{Schmoll_2019}.

\subsubsection*{Zero-flux limit}

Another insightful case is the zero-flux case, \({f_{ijk}^{+} = f_{ijk}^{-} = 0}\), but with non-zero asymmetry in hopping amplitudes, \({\lambda \neq 1}\). In this case the phase factors \({\theta^{-}_{ij} = 0}\) and \({h_{ijk}^{-} \equiv h^{-}}\), \({w_{ijk}^{-} \equiv w^{-}}\) are independent of the indices. 
The effective Hamiltonian then reads
\begin{align}
\label{eq:H_eff_Spin_zero_flux}
	\mathcal{H}_{\mathrm{eff}}
	&=
	\sum_{i < j}
		J_{ij}
		\Big[
			\frac{1}{2}
			\left(
				S^{+}_{i}
				S^{-}_{j}
				+
				\mathrm{H.c.}
			\right)
			+
			\gamma
			S^{z}_{i}
			S^{z}_{j}
		\Big]
\nonumber
\\
	&\mathrel{\hphantom{=}}
	\mathrel{+}
	\sum_{i < j < k}
		J_{ijk}	
		\bigg\{
			h^{-}
			\left(S^{z}_i + S^{z}_j + S^{z}_k-12S^{z}_i S^{z}_j S^{z}_k\right)
\nonumber
\\
			&\mathrel{\hphantom{=}} \qquad\qquad\qquad
			\mathrel{+}
			12
			w^{-}
			\Big[
				\left(
                    \bm{S}^{\vphantom{z}}_i
                    \cdot
                    \bm{S}^{\vphantom{z}}_j
                \right)
                S^z_k
				+
				\left(
                    \bm{S}^{\vphantom{z}}_j 
                    \cdot 
                    \bm{S}^{\vphantom{z}}_k
                \right)
                S^z_i
				+
				\left(
                    \bm{S}^{\vphantom{z}}_k 
                    \cdot 
                    \bm{S}^{\vphantom{z}}_i
                \right)
                S^z_j
				-
				3
				S^{z}_i
				S^{z}_j
				S^{z}_k
			\Big]
		\bigg\}
\,,
\end{align}
where
\begin{align}
	h^{-} = \lambda^3 - \lambda^{-3} \,,\quad
	w^{-} = \lambda^{-1} - \lambda \,.
\end{align}

\subsubsection*{Falicov-Kimball limit}
In the Falicov-Kimball model~\cite{FK_69} particles with one spin projection, e.g.\ minus-spin particles are immobile, \({t_{x}^{-} = 0}\). 
In this case, the quantity ${\lambda \to \infty}$ is ill-defined, however, the products $J_{ijk} \lambda$ and ${J_{ijk} \lambda^3}$ are well-defined. For that matter, it is more instructive to use the expressions for the hopping terms written explicitly in terms of spin-dependent hopping amplitudes $t_x^{\sigma}$. There it is clear that $\mathcal{H}_{\mathrm{eff}}$ reduces to Ising-type model, in particular
\begin{align}
\label{eq:H'_2_FK}
	\mathcal{H}_{\mathrm{eff}}
	=
	\sum_{i < j}
		J_{ij}^{\vphantom{z}}
		S^{z}_{i}
		S^{z}_{j}
		+
	\sum_{i < j < k}
		J_{ijk}^{\vphantom{-}}
		h_{ijk}^{-}
		\left(
			S^{z}_i + S^{z}_j + S^{z}_k
			-
			12
			S^{z}_i S^{z}_j S^{z}_k
		\right)
,
\end{align}
where
\begin{align}
\qquad
	J_{ij}
	=
	\frac{2\big|t_{ij}^{+}\big|^2}{U}
\,,
\quad  
    J_{ijk}
    =
    \frac{\big|t_{ij}^{+} t_{jk}^{+} t_{ki}^{+}\big|}{U^2}
\,,
\quad
    h_{ijk}^{-}
    =
    \cos (f_{ijk}^{+})
\,.
\end{align}

The modified Ising model, which includes the extended three-spin magnetic field term, has been exactly solved in Ref.~\cite{ising_3spin}. There, using the transfer matrix method, the statistical sum was computed, and the magnetization in the ground state has been obtained. 
However, in the context of ultracold atoms in optical lattices, the magnetization of the system is determined by the population difference between the two species of atoms, which is not affected by the variation of the external parameters. Therefore, physically relevant formulation of the problem is to find the ground state configuration for the fixed magnetization.

\section{The strong attractive interaction \texorpdfstring{$U<0$}{U<0}} 
\label{sec:Effective_Ham_U<0}

\subsection{Particle-hole transformation}
\label{sec:PH_transformation}

In order to obtain the effective Hamiltonian for \({U<0}\), similar to~\eqref{eq:H_eff_Spin}, one could repeat the same procedure as in Sec.~\ref{sec:H_eff_T} and App.~\ref{sec:App_H_eff_derivation}. 
However, in the case of strong attractive interaction, the lowest energy state has only empty or doubly occupied sites.
This fact significantly alters the technical details, requiring all the tedious calculations in Sec.~\ref{sec:H_eff_T} and App.~\ref{sec:App_H_eff_derivation} to be rederived from scratch.
For that matter, we use a particle-hole transformation to construct the effective Hamiltonian for \({U<0}\) in a simple way from the already obtained one for \({U>0}\), Eq.~\eqref{eq:H_eff_Spin} (consult e.g.\ \cite{1D_Hubbard_book}).
For clarity, we perform a particle-hole transformation for \({\sigma=-}\) particles, which is defined by
\begin{align}
\label{eq:ph_transformation_down}
	c^{\vphantom{\dag}}_{i -}
    =
	\tilde{c}^\dag_{i -}
\,,
\end{align}
in the sense of mapping \({n_{i-} = \tilde{h}_{i-}}\) and \({h_{i-} = \tilde{n}_{i-}}\). As a result, the number of new \({\sigma = -}\) particles becomes
\begin{align}
\label{eq:N_down_transform}
	\tilde{N}_{-}
	=
	N - N_{-}
\,.
\end{align}
In case of zero net magnetization at half-filling, we have \({\tilde{N}_- = N_-}\).
Thus, the half-filling and zero net magnetization conditions are preserved.

Leaving \({\sigma=+}\) particles intact, \({c^{\vphantom{\dag}}_{i +} = \tilde{c}^{\vphantom{\dag}}_{i +}}\), the local Hilbert spaces, corresponding to transformed $\tilde{c}^{\vphantom{\dag}}_{i\sigma}, \tilde{c}^\dag_{i\sigma}$ and untransformed $c^{\vphantom{\dag}}_{i\sigma}, c^\dag_{i\sigma}$ operators are related as
\begin{align}
\label{eq:Hilbert_space_transform}
	| 0 \rangle = | \tilde{-} \rangle \,,\qquad
	| {-} \rangle = | \tilde{0} \rangle \,,\qquad
	| {+} \rangle = | \tilde{2} \rangle \,,\qquad
	| 2 \rangle = | \tilde{+} \rangle \,.
\end{align}
The relation~\eqref{eq:Hilbert_space_transform} describes a reinterpretation of the physical states. The Hilbert spaces spanned by the states $|a\rangle$ and $|\tilde{a}\rangle$ are the same, where \({a \in \{0,+,-,2\} }\) and \({\tilde{a} \in \{\tilde{0}, \tilde{+}, \tilde{-}, \tilde{2}\} }\).
The Hamiltonian~\eqref{eq:H_ij} under the correspondence~\eqref{eq:ph_transformation_down} transforms as
\begin{align}
\label{eq:H_transform}
	\widetilde{H}
	=
	\widetilde{T}
	+
	\widetilde{V}
	=
	-
	\sum_{ij}
		t_{ij}^{+}
		\tilde{c}^\dag_{i+}
		\tilde{c}^{\vphantom{\dag}}_{j+}
	+
	\sum_{ij}
		\left(t_{ij}^{-}\right)^{\! *}
		\tilde{c}^\dag_{i-}
		\tilde{c}^{\vphantom{\dag}}_{j-}
	-
	\frac{U}{2}
	\sum_{i \sigma}
		\tilde{n}_{i \sigma}
		\tilde{n}_{i \bar{\sigma}}
	+
	U
	\sum_{i}
		\tilde{n}_{i+}
\,.
\end{align}
Neglecting the irrelevant constant \({U N_{+}}\), the transformed Hamiltonian~\eqref{eq:H_transform} returns to the original form~\eqref{eq:H_ij} by simultaneous substitutions
\begin{subequations}
\label{eq:t_U_tilde}
\begin{align}
\label{eq:U_tilde}
	U &\to -U
\,,
\\
\label{eq:t_varphi_tilde}
	\left(t_{ij}^{-}\right)^{\! *}
	&\to
	-t_{ij}^{-}
\iff
	\varphi_{ij}^{-}
	\to
	-
	\varphi_{ij}^{-}
	+
	\pi
\,.
\end{align}
\end{subequations}
In other words, the following identity holds up to the irrelevant additive constant
\begin{align}
\label{eq:H_identity}
    H\!\left(\varphi_{ij}^{-}, U;\, c^{\vphantom{\dag}}_{i\sigma}, c^\dag_{i\sigma}\right)
    =
    H\!\left(-\varphi_{ij}^{-} + \pi, -U;\, \tilde{c}^{\vphantom{\dag}}_{i\sigma}, \tilde{c}^\dag_{i\sigma}\right)
    +
    \mathrm{const}
\,.
\end{align}
Here the arguments $|t_{ij}^\sigma|$ and $\varphi_{ij}^{+}$, that are the same on both sides, are not indicated for brevity. The functional identity~\eqref{eq:H_identity} is an exact one, not related to any approximation scheme.

To summarize, the prescription is as follows. Obtain effective spin \({S = 1/2}\) Hamiltonian for \({U > 0}\), i.e.\ Eq.~\eqref{eq:H_eff_Spin}. Make substitutions~\eqref{eq:t_U_tilde}. Neglecting the constant \({U N_{+}}\), the result will be valid for \({U < 0}\), but in the sense of reinterpreted Hilbert space~\eqref{eq:Hilbert_space_transform}. This means that spin ${S=1/2}$ operators should be swapped with pseudospin (charge) ${\eta=1/2}$ operators.
\begin{align}
\label{eq:H_eff_Casrge}
	U<0:
\quad
	\mathcal{H}_{\mathrm{eff}}
	=
	&-
	\sum_{i < j}
		J_{ij}
		\Big[
			\frac{1}{2}
			\left(
				e^{\mathrm{i} \theta^{+}_{ij}}
				\eta^{+}_{i}
				\eta^{-}_{j}
				+
				\mathrm{H.c.}
			\right)
			-
			\gamma
			\eta^{z}_{i}
			\eta^{z}_{j}
		\Big]
\nonumber
\\
	&+
	\sum_{i < j < k}
		J_{ijk}^{\vphantom{+}}
		\bigg\{
			h_{ijk}^{+}
			\left(\eta^{z}_i + \eta^{z}_j + \eta^{z}_k-12\eta^{z}_i \eta^{z}_j \eta^{z}_k\right)
\nonumber
\\
			&\mathrel{\hphantom{=}} \qquad
			\mathrel{+}
			6
			\Big[
				w_{ijk}^{+}
				\left(
					e^{\mathrm{i} \theta^{+}_{ij}}
					\eta^{+}_i
					\eta^{-}_j
					\eta^{z}_k
					+
					e^{\mathrm{i} \theta^{+}_{jk}}
					\eta^{+}_j
					\eta^{-}_k
					\eta^{z}_i 
					+
					e^{\mathrm{i} \theta^{+}_{ki}}
					\eta^{+}_k
					\eta^{-}_i
					\eta^{z}_j
				\right)
				+
				\mathrm{H.c.}
			\Big]
		\bigg\}
\,,
\end{align}
where
\begin{align}
\label{eq:J_ij_U<0}
	J_{ij}
	&=
	\frac{4\tau_{ij}^2}{|U|}
\,,
\qquad
	\gamma
	=
	\left(\lambda^{2} + \lambda^{-2}\right) \! / 2
\,,
\quad
	\left(\gamma \geqslant 1\right)
,
\\
\label{eq:J_ijk_U<0}
	J_{ijk}
	&=
	\frac{\tau_{ij} \tau_{jk} \tau_{ki}}{U^2}
\,,
\\
\label{eq:theta_ij_U<0}
	\theta^{+}_{ij}
	&=
	\varphi_{ij}^{+} + \varphi_{ij}^{-}
\,,
\\
\label{eq:h_ijk_U<0}
	h_{ijk}^{+}
	&=
	\lambda^{3}
	\cos \! \left(f_{ijk}^{+}\right)
	+
	\lambda^{-3}
	\cos \! \left(f_{ijk}^{-}\right)
,
\\
\label{eq:w_ijk_U<0}
	w_{ijk}^{+}
	&=
	\lambda^{-1}
	\exp \! \left(\mathrm{i} f_{ijk}^{-}\right)
	+
	\lambda
	\exp \! \left(\mathrm{i} f_{ijk}^{+}\right)
.
\end{align}
Therefore, the limiting cases considered in the Sec.~\ref{sec:limits} can be easily reproduced for ${U<0}$, now in terms of the pseudospin (charge) $\eta^{\pm}$ and $\eta^{z}$ operators.

Before we close the section, we highlight the following fact. 
For \({U > 0}\), in the spin-symmetric case, i.e.\ \({\lambda = 1}\) and \({f_{ijk}^{+} = f_{ijk}^{-}}\), the DMI and the magnetic field vanish. The corresponding condition for \({U < 0}\) looks like \({f_{ijk}^{+} = -f_{ijk}^{-} + \pi}\), due to transformation~\eqref{eq:t_varphi_tilde}. In general, these two conditions are not satisfied simultaneously and thus, if the DMI and the magnetic field vanish in one sector, they are still present in the other. The only exception is the particle-hole symmetric case, where fluxes have a value \({|f_{ijk}^{\sigma}| = \pi/2}\). In this case, the DMI and the magnetic field are absent in both sectors, \({U>0}\) and \({U<0}\).

\section{Conclusion}
\label{sec:Summary}

In this paper we have derived the effective spin and pseudospin (charge) Hamiltonians, describing the infrared properties of the one-dimensional half-filled spin-asymmetric Hubbard model on a triangular ladder with spin-dependent flux penetrating the lattice, corresponding to the spin and charge sectors in the limit of strong on-site repulsion \({(U>0)}\) and attraction \({(U<0)}\), respectively.
We showed that for \({U>0}\) the obtained Hamiltonian is that of anisotropic $XXZ$ Heisenberg zig-zag ladder with the DMI and a three-spin coupling.
Due to the presence of spin-dependent flux, the coupling constants acquire phase factors $e^{\mathrm{i} \theta^{-}_{ij}}$ and are complex valued. The real part of the coupling in the second order Hamiltonian $\mathcal{H}_{\mathrm{eff}}^{(2)}$ constitutes a usual anisotropic $XXZ$ Heisenberg model, while the imaginary part corresponds to the DMI with orientation vector along $z$-axis, describing a spin current in terms of original fermions. 
Spin-dependent hopping amplitudes induce restricted exchange anisotropy \({\gamma \geqslant 1}\) in the system. The three-spin part of the Hamiltonian  consists of two kinds of terms. The first one corresponds to the oriented in $z$-direction magnetic field term but modified by the additional three-spin Ising term given by the product of spin $z$-components located on vertices of each triangular plaquette. The second term describes a correlated spin exchange (CSE) with complex amplitudes. In case of spin-symmetric hopping and flux, the magnetic field and the Dzyaloshinskii-Moriya terms vanish, while the CSE part reduces to scalar spin chirality operator.

We have used the one-spin-component particle-hole transformation to map the repulsive ${(U>0)}$ case onto the attractive ${(U<0)}$ one and to construct an effective Hamiltonian in the limit of strong attraction. The obtained Hamiltonian describes the charge degrees of freedom in terms of doubly occupied sites (doublons) and empty sites (holons). Their dynamics is again described by the same $XXZ$ Heisenberg model, but in terms of pseudospin (charge) operators.
The magnetic field and the Dzyaloshinskii-Moriya terms vanish in the limit of anti-symmetric flux (modulo $\pi$) configuration. 
In the latter case, the three-spin coupling term is given by the scalar pseudospin chirality operator.

\section{Acknowledgments}

The authors would like to thank Alexander A. Nersesyan for stimulating 
interest in this work with numerous enlightening discussions.

\appendix
\section{Derivation of effective Hamiltonian}
\label{sec:App_H_eff_derivation}

In this appendix we provide the technical details of the necessary calculations to obtain the effective Hamiltonian~\eqref{eq:H_eff_Spin}, which we skipped in the main text.
\subsection{\texorpdfstring{$\mathcal{H}_{\mathrm{eff}}^{(2)}$}{H'2}}
Computation of $\mathcal{H}_{\mathrm{eff}}^{(2)}$ starts with plugging Eqs.~\eqref{eq:T+_to_X} and \eqref{eq:T-_to_X} into Eq.~\eqref{eq:H'_2}
\begin{subequations}
\begin{align}
	T_{-1}
	T_{1}
	&=
	\sum_{i_{1} j_{1} \sigma_{1}}
	\sum_{i_{2} j_{2} \sigma_{2}}
		\sigma_{1}
		\sigma_{2}
		t_{i_{1} j_{1}}^{\sigma_{1}}
		t_{i_{2} j_{2}}^{\sigma_{2}}
		X^{\sigma_{1} 0}_{i_{1}}
		X^{\bar{\sigma}_{1} 2}_{j_{1}}
		X^{2 \bar{\sigma}_{2}}_{i_{2}}
		X^{0 \sigma_{2}}_{j_{2}}
\,;
\intertext{
The product of $X$ operators is non-zero only for \({i_{1} = j_{2} \equiv i}\), \({i_{2} = j_{1} \equiv j}\). In that case they contract according to Eq.~\eqref{eq:X_contraction} and commute or anticommute according to Eqs.~\eqref{eq:X_comm_anticomm_rules}
}
	T_{-1}
	T_{1}
	&=
	\sum_{i j}
	\sum_{\sigma_{1} \sigma_{2}}
		\sigma_{1}
		\sigma_{2}
		t_{ij}^{\sigma_{1}}
		t_{ji}^{\sigma_{2}}
		X^{\sigma_{1} \sigma_{2}}_{i}
		X^{\bar{\sigma}_{1} \bar{\sigma}_{2}}_{j}
\,;
\intertext{
Splitting the summation over spin indices into two parts, one with \({\sigma_{1} = \sigma_{2} \equiv \sigma}\) and another with \({\sigma_{1} = \bar{\sigma}_{2} \equiv \sigma}\), we obtain
}
	T_{-1}
	T_{1}
	&=
	\sum_{i j \sigma}
	\left(
		\big|t_{ij}^{\sigma}\big|^{2}
		X^{\sigma \sigma}_{i}
		X^{\bar{\sigma} \bar{\sigma}}_{j}
		-
		t_{ij}^{\sigma}
		\left(t_{ij}^{\bar{\sigma}}\right)^{\! *}
		X^{\sigma \bar{\sigma}}_{i}
		X^{\bar{\sigma} \sigma}_{j}
	\right)
;
\intertext{
Plugging the correspondence~\eqref{eq:X-to-S} yields
}
\label{eq:T_{-1}T_{1}_in_terms_of_t}
	T_{-1}
	T_{1}
	&=
	\sum_{i j \sigma}
	\left[
		\big|t_{ij}^{\sigma}\big|^{2}
		\bigg(\frac{1}{2} + \sigma S^{z}_{i}\bigg)
		\bigg(\frac{1}{2} - \sigma S^{z}_{j}\bigg)
		-
		t_{ij}^{\sigma}
		\left(t_{ij}^{\bar{\sigma}}\right)^{\! *}
		S^{\sigma}_{i}
		S^{\bar{\sigma}}_{j}
	\right]
;
\end{align}
\end{subequations}
Noting that in the last term, \({\sum_{\sigma} t_{ij}^{\sigma} \left(t_{ij}^{\bar{\sigma}}\right)^{\! *} S^{\sigma}_{i} S^{\bar{\sigma}}_{j}}\), Hermitian conjugation, spin flip \({\sigma \leftrightarrow \bar\sigma}\) and site exchange \({i \leftrightarrow j}\) are three equivalent operations, we arrive at the final form
\begin{align}
\label{eq:H'_2_effective_S^pm}
	\mathcal{H}_{\mathrm{eff}}^{(2)}
	=
	\sum_{i < j}
		J_{ij}
		\bigg[
			\frac{1}{2}
			\left(
				e^{\mathrm{i} \theta^{-}_{ij}}
				S^{+}_{i}
				S^{-}_{j}
				+
				\mathrm{H.c.}
			\right)
			+
			\gamma
			S^{z}_{i}
			S^{z}_{j}
		\bigg]
,
\end{align}
with
\begin{align}
\label{eq:J_ij_2}
	J_{ij}
	&=
	4 \tau_{ij}^2 / U
\,,
\\
\label{eq:gamma}
	\gamma
	&=
	\left(\lambda^{2} + \lambda^{-2}\right) \! / 2
\,,
\quad
	\left(\gamma \geqslant 1\right)
,
\end{align}
where \({\tau_{ij} = |t_{ij}^{+} t_{ij}^{-}|^{1/2}}\), \({\lambda = |t_{ij}^{+} / t_{ij}^{-}|^{1/2}}\) (independent of $i,j$ indices) and \({\theta^{-}_{ij} = \varphi_{ij}^{+} - \varphi_{ij}^{-}}\).

\subsection{\texorpdfstring{$\mathcal{H}_{\mathrm{eff}}^{(3)}$}{H'3}}
Similar to $\mathcal{H}_{\mathrm{eff}}^{(2)}$, computation of $\mathcal{H}_{\mathrm{eff}}^{(3)}$ starts with plugging Eqs.~\eqref{eq:T_to_X} in Eq.~\eqref{eq:H'_3}
\begin{subequations}
\begin{align}
	T_{-1}
	T_{0}
	T_{1}
	&=
	-
	\sum_{i_{1} j_{1} \sigma_{1}}
	\sum_{i_{2} j_{2} \sigma_{2}}
	\sum_{i_{3} j_{3} \sigma_{3}}
		\sigma_{1}
		\sigma_{3}
		\,
		t_{i_{1} j_{1}}^{\sigma_{1}}
		t_{i_{2} j_{2}}^{\sigma_{2}}
		t_{i_{3} j_{3}}^{\sigma_{3}}
		X^{\sigma_{1} 0}_{i_{1}}
		X^{\bar{\sigma}_{1} 2}_{j_{1}}
		\left(
			X^{2 \bar{\sigma}_{2}}_{i_{2}}
			X^{\bar{\sigma}_{2} 2}_{j_{2}}
			+
			X^{\sigma_{2} 0}_{i_{2}}
			X^{0 \sigma_{2}}_{j_{2}}
		\right)
		X^{2 \bar{\sigma}_{3}}_{i_{3}}
		X^{0 \sigma_{3}}_{j_{3}}
\nonumber
\\
	&=
	-
	\sum_{i_{1} j_{1} \sigma_{1}}
	\sum_{i_{2} j_{2} \sigma_{2}}
	\sum_{i_{3} j_{3} \sigma_{3}}
		\sigma_{1}
		\sigma_{3}
		\,
		t_{i_{1} j_{1}}^{\sigma_{1}}
		t_{i_{2} j_{2}}^{\sigma_{2}}
		t_{i_{3} j_{3}}^{\sigma_{3}}
		X^{\sigma_{1} 0}_{i_{1}}
		X^{\bar{\sigma}_{1} 2}_{j_{1}}
		X^{2 \bar{\sigma}_{2}}_{i_{2}}
		X^{\bar{\sigma}_{2} 2}_{j_{2}}
		X^{2 \bar{\sigma}_{3}}_{i_{3}}
		X^{0 \sigma_{3}}_{j_{3}}
\nonumber
\\
	&\mathrel{\hphantom{=}}
	-
	\sum_{i_{1} j_{1} \sigma_{1}}
	\sum_{i_{2} j_{2} \sigma_{2}}
	\sum_{i_{3} j_{3} \sigma_{3}}
		\sigma_{1}
		\sigma_{3}
		\,
		t_{i_{1} j_{1}}^{\sigma_{1}}
		t_{i_{2} j_{2}}^{\sigma_{2}}
		t_{i_{3} j_{3}}^{\sigma_{3}}
		X^{\sigma_{1} 0}_{i_{1}}
		X^{\bar{\sigma}_{1} 2}_{j_{1}}
		X^{\sigma_{2} 0}_{i_{2}}
		X^{0 \sigma_{2}}_{j_{2}}
		X^{2 \bar{\sigma}_{3}}_{i_{3}}
		X^{0 \sigma_{3}}_{j_{3}}
\,;
\intertext{
In the first sum, the product of $X$ operators is non-zero only for \({i_{1} = j_{3} \equiv i}\), \({i_{2} = j_{1} \equiv j}\), \({i_{3} = j_{2} \equiv k}\), while in the second sum for \({j_{1} = i_{3} \equiv i}\), \({j_{2} = i_{1} \equiv j}\), \({j_{3} = i_{2} \equiv k}\). In that case they contract according to Eq.~\eqref{eq:X_contraction} and commute or anticommute according to Eqs.~\eqref{eq:X_comm_anticomm_rules}
}
	T_{-1}
	T_{0}
	T_{1}
	&=
	-
	\sum_{i j k}
	\sum_{\sigma_{1} \sigma_{2} \sigma_{3}}
		\sigma_{1}
		\sigma_{3}
		\,
		t_{ij}^{\sigma_{1}}
		t_{jk}^{\sigma_{2}}
		t_{ki}^{\sigma_{3}}
		X^{\sigma_{1} \sigma_{3}}_{i}
		X^{\bar{\sigma}_{1} \bar{\sigma}_{2}}_{j}
		X^{\bar{\sigma}_{2} \bar{\sigma}_{3}}_{k}
\nonumber
\\
	&\mathrel{\hphantom{=}}
	+
	\sum_{i j k}
	\sum_{\sigma_{1} \sigma_{2} \sigma_{3}}
		\sigma_{1}
		\sigma_{3}
		\left(
			t_{ij}^{\sigma_{1}}
			t_{jk}^{\sigma_{2}}
			t_{ki}^{\sigma_{3}}
		\right)^{\! *}
		X^{\bar{\sigma}_{1} \bar{\sigma}_{3}}_{i}
		X^{\sigma_{1} \sigma_{2}}_{j}
		X^{\sigma_{2} \sigma_{3}}_{k}
\,;
\intertext{Renaming the spin indices in the second sum \({\sigma_1, \sigma_2, \sigma_3 \to \bar{\sigma}_1, \bar{\sigma}_2, \bar{\sigma}_3}\) we obtain
}
	T_{-1}
	T_{0}
	T_{1}
	&=
	-
	\sum_{i j k}
	\sum_{\sigma_{1} \sigma_{2} \sigma_{3}}
		\sigma_{1}
		\sigma_{3}
		\left[
			t_{ij}^{\sigma_{1}}
			t_{jk}^{\sigma_{2}}
			t_{ki}^{\sigma_{3}}
			-
			\left(
				t_{ij}^{\bar{\sigma}_{1}}
				t_{jk}^{\bar{\sigma}_{2}}
				t_{ki}^{\bar{\sigma}_{3}}
			\right)^{\! *}
		\,\right]
		X^{\sigma_{1} \sigma_{3}}_{i}
		X^{\bar{\sigma}_{1} \bar{\sigma}_{2}}_{j}
		X^{\bar{\sigma}_{2} \bar{\sigma}_{3}}_{k}
\,;
\intertext{
In the sum over spin indices, there are four cases: 1)~\({\sigma_1 = \sigma_2 = \sigma_3 \equiv \sigma}\), 2)~\({\sigma_1 \equiv \bar{\sigma}, \, \sigma_2 = \sigma_3 \equiv \sigma}\), 3)~\({\sigma_1 = \sigma_3 \equiv \sigma}\), \ \({\sigma_2 \equiv \bar{\sigma}}\), 4)~\({\sigma_3 \equiv \bar{\sigma}, \, \sigma_1 = \sigma_2 \equiv \sigma}\). Splitting the sum accordingly into four parts we obtain
}
	T_{-1}
	T_{0}
	T_{1}
	&=
	-
	\sum_{i j k \sigma}
		\left[
			t_{ij}^{\sigma}
			t_{jk}^{\sigma}
			t_{ki}^{\sigma}
			-
			\left(
				t_{ij}^{\bar{\sigma}}
				t_{jk}^{\bar{\sigma}}
				t_{ki}^{\bar{\sigma}}
			\right)^{\! *}
		\,\right]
		X^{\sigma \sigma}_{i}
		X^{\bar{\sigma} \bar{\sigma}}_{j}
		X^{\bar{\sigma} \bar{\sigma}}_{k}
\nonumber
\\
	&\mathrel{\hphantom{=}}
	+
	\sum_{i j k \sigma}
		\left[
			t_{ij}^{\bar{\sigma}}
			t_{jk}^{\sigma}
			t_{ki}^{\sigma}
			-
			\left(
				t_{ij}^{\sigma}
				t_{jk}^{\bar{\sigma}}
				t_{ki}^{\bar{\sigma}}
			\right)^{\! *}
		\,\right]
		X^{\bar{\sigma} \sigma}_{i}
		X^{\sigma \bar{\sigma}}_{j}
		X^{\bar{\sigma} \bar{\sigma}}_{k}
\nonumber
\\
	&\mathrel{\hphantom{=}}
	-
	\sum_{i j k \sigma}
		\left[
			t_{ij}^{\sigma}
			t_{jk}^{\bar{\sigma}}
			t_{ki}^{\sigma}
			-
			\left(
				t_{ij}^{\bar{\sigma}}
				t_{jk}^{\sigma}
				t_{ki}^{\bar{\sigma}}
			\right)^{\! *}
		\,\right]
		X^{\sigma \sigma}_{i}
		X^{\bar{\sigma} \sigma}_{j}
		X^{\sigma \bar{\sigma}}_{k}
\nonumber
\\
	&\mathrel{\hphantom{=}}
	+
	\sum_{i j k \sigma}
		\left[
			t_{ij}^{\sigma}
			t_{jk}^{\sigma}
			t_{ki}^{\bar{\sigma}}
			-
			\left(
				t_{ij}^{\bar{\sigma}}
				t_{jk}^{\bar{\sigma}}
				t_{ki}^{\sigma}
			\right)^{\! *}
		\,\right]
		X^{\sigma \bar{\sigma}}_{i}
		X^{\bar{\sigma} \bar{\sigma}}_{j}
		X^{\bar{\sigma} \sigma}_{k}
\,;
\intertext{
Plugging the correspondence~\eqref{eq:X-to-S} yields
}
	T_{-1}
	T_{0}
	T_{1}
	&=
	-
	\sum_{i j k \sigma}
		\left[
			t_{ij}^{\sigma}
			t_{jk}^{\sigma}
			t_{ki}^{\sigma}
			-
			\left(
				t_{ij}^{\bar{\sigma}}
				t_{jk}^{\bar{\sigma}}
				t_{ki}^{\bar{\sigma}}
			\right)^{\! *}
		\,\right]
		\left(\frac{1}{2} + \sigma S^{z}_{i}\right)
		\left(\frac{1}{2} - \sigma S^{z}_{j}\right)
		\left(\frac{1}{2} - \sigma S^{z}_{k}\right)
\nonumber
\\
	&\mathrel{\hphantom{=}}
	+
	\sum_{i j k \sigma}
		\left[
			t_{ij}^{\bar{\sigma}}
			t_{jk}^{\sigma}
			t_{ki}^{\sigma}
			-
			\left(
				t_{ij}^{\sigma}
				t_{jk}^{\bar{\sigma}}
				t_{ki}^{\bar{\sigma}}
			\right)^{\! *}
		\,\right]
		S^{\bar{\sigma}}_{i}
		S^{\sigma}_{j}
		\left(\frac{1}{2} - \sigma S^{z}_{k}\right)
\nonumber
\\
	&\mathrel{\hphantom{=}}
	-
	\sum_{i j k \sigma}
		\left[
			t_{ij}^{\sigma}
			t_{jk}^{\bar{\sigma}}
			t_{ki}^{\sigma}
			-
			\left(
				t_{ij}^{\bar{\sigma}}
				t_{jk}^{\sigma}
				t_{ki}^{\bar{\sigma}}
			\right)^{\! *}
		\,\right]
		\left(\frac{1}{2} + \sigma S^{z}_{i}\right)
		S^{\bar{\sigma}}_{j}
		S^{\sigma}_{k}
\nonumber
\\
\label{eq:T_{-1}T_{0}T_{1}_in_terms_of_t}
	&\mathrel{\hphantom{=}}
	+
	\sum_{i j k \sigma}
		\left[
			t_{ij}^{\sigma}
			t_{jk}^{\sigma}
			t_{ki}^{\bar{\sigma}}
			-
			\left(
				t_{ij}^{\bar{\sigma}}
				t_{jk}^{\bar{\sigma}}
				t_{ki}^{\sigma}
			\right)^{\! *}
		\,\right]
		S^{\sigma}_{i}
		\left(\frac{1}{2} - \sigma S^{z}_{j}\right)
		S^{\bar{\sigma}}_{k}
\,;
\end{align}
\end{subequations}
Summing over $\sigma$ and all permutations of site indices in a given triangle, the terms containing even number of $S$ operators gets canceled. After tedious but straightforward algebra we arrive at the final form
\begin{align}
\label{eq:H'_3_general}
	\mathcal{H}_{\mathrm{eff}}^{(3)}
	\mathrel{=}
	&\sum_{i < j < k}
	J_{ijk}
	\bigg\{
	h_{ijk}^{-}
	\left(
		S^{z}_i + S^{z}_j + S^{z}_k
		-
		12
		S^{z}_i S^{z}_j S^{z}_k
	\right)
\nonumber
\\
	&\qquad\qquad\qquad
	+
	6
	\left[
		w_{ijk}^{-}
		\left(
			e^{\mathrm{i} \theta^{-}_{ij}}
			S^{+}_i
			S^{-}_j
			S^{z}_k
			+
			e^{\mathrm{i} \theta^{-}_{jk}}
			S^{+}_j
			S^{-}_k
			S^{z}_i
			+
			e^{\mathrm{i} \theta^{-}_{ki}}
			S^{+}_k
			S^{-}_i
			S^{z}_j
		\right)
		+
		\mathrm{H.c.}
	\right]
	\!\bigg\}
\,,
\end{align}
with
\begin{align}
	J_{ijk}
	&=
	\frac{\tau_{ij} \tau_{jk} \tau_{ki}}{U^2}
\,,
\\
	h_{ijk}^{-}
	&=
	\lambda^{3}
	\cos \! \left(f_{ijk}^{+}\right)
	-
	\lambda^{-3}
	\cos \! \left(f_{ijk}^{-}\right)
,
\\
	w_{ijk}^{-}
	&=
	\lambda^{-1}
	\exp \! \left(-\mathrm{i} f_{ijk}^{-}\right)
	-
	\lambda
	\exp \! \left(\mathrm{i} f_{ijk}^{+}\right)
,
\end{align}
where $f_{ijk}^{\sigma}$ is equal to either $f_{A}^{\sigma}$ or $-f_{B}^{\sigma}$, depending on which triangle the indices $i,j,k$ (with \({i<j<k}\)) form.

\section{Local gauge transformation}
\label{sec:App_gauge}
The fluxes \({f_{A}^{\sigma}}\) and \({f_{B}^{\sigma}}\) are physical observables, however, individual phases \({\varphi_{x}^{\sigma}}\) are not uniquely defined.
In this appendix, we discuss two gauge choices, which are convenient for different purposes. 

For each $\sigma$, we consider the local gauge transformation of the form
\begin{subequations}
\label{eq:U(1)_general}
\begin{align}
	a_{\ell \sigma}
	&\to
	e^{\mathrm{i}\left(\alpha_{\sigma} \ell - \beta_{\sigma}/2\right)}
	a_{\ell \sigma}
\,,
\\
	b_{\ell \sigma}
	&\to
	e^{\mathrm{i}\left(\alpha_{\sigma} \ell + \beta_{\sigma}/2\right)}
	b_{\ell \sigma}
\,.
\end{align}
\end{subequations}
Note that taking different \({\alpha_\sigma}\) for $A$ and $B$ would lead to $\ell$-dependent $\varphi_{x}^{\sigma}$ phases in Hamiltonian~\eqref{eq:H_AB}, while taking different \({\beta_\sigma}\) for $A$ and $B$ would be redundant, as only their difference would appear in the terms of the type $a^\dag b$.
The transformation~\eqref{eq:U(1)_general} maps the $\varphi_{x}^{\sigma}$ phases of Hamiltonian~\eqref{eq:H_AB} to
\begin{subequations}
\label{eq:transformed_phases}
\begin{alignat}{4}
	&
	\varphi_{1}^{\sigma}
	&&
	\to
	\varphi_{1}^{\sigma}
	-
	\beta_{\sigma}
\,,
\qquad
	&&
	\varphi_{2}^{\sigma}
	&&
	\to
	\varphi_{2}^{\sigma}
	-
	\alpha_{\sigma}
	+
	\beta_{\sigma}
\,,
\\
	&
	\varphi_{A}^{\sigma}
	&&
	\to
	\varphi_{A}^{\sigma}
	+
	\alpha_{\sigma}
\,,
\qquad
	&&
	\varphi_{B}^{\sigma}
	&&
	\to
	\varphi_{B}^{\sigma}
	+
	\alpha_{\sigma}
\,.
\end{alignat}
\end{subequations}
Obviously, the fluxes in $A$- and $B$-triangles~\eqref{eq:fluxes} are gauge invariant.

\subsection{Gauge with \texorpdfstring{$\varphi_1^\pm = \varphi_2^\pm = 0$}{vaprhi-1,2,+- = 0}}
\label{sec:App_gauge_corollary}

The hopping parameters along a zig-zag line can be made real, i.e.\ \({\varphi_{1}^{\sigma}, \varphi_{2}^{\sigma} \to 0}\), by choosing the parameters
\begin{subequations}
\begin{align}
	\alpha_{\sigma}
	&=
	\varphi_{1}^{\sigma}
	+
	\varphi_{2}^{\sigma}
\,,
\\
	\beta_{\sigma}
	&=
	\varphi_{1}^{\sigma}
\,.
\end{align}
\end{subequations}
According to Eq.~\eqref{eq:transformed_phases} the intra-chain hopping phases transform to
\begin{subequations}
\begin{align}
	\varphi_{A}^{\sigma}
	&\to
	\varphi_{A}^{\sigma}
	+
	\varphi_{1}^{\sigma}
	+
	\varphi_{2}^{\sigma}
	=
	f_{A}^{\sigma}
\,,
\\
	\varphi_{B}^{\sigma}
	&\to
	\varphi_{B}^{\sigma}
	+
	\varphi_{1}^{\sigma}
	+
	\varphi_{2}^{\sigma}
	=
	-f_{B}^{\sigma}
\,,
\end{align}
\end{subequations}
which are the gauge invariant fluxes in $A$- and $B$-triangles. 

Corollary is that there exists a gauge with \({\theta^{-}_{x} = \varphi_{x}^{+} - \varphi_{x}^{-} = 0 \mod{\pi}}\) for all \({x \in \{A, B, 1, 2\}}\) if and only if fluxes are spin-symmetric modulo $\pi$, i.e.\ \({f_{A}^{+} = f_{A}^{-} \mod{\pi}}\) and \({f_{B}^{+} = f_{B}^{-} \mod{\pi}}\). 
Similarly, there exists a gauge with \({\theta^{+}_{x} = \varphi_{x}^{+} + \varphi_{x}^{-} = 0 \mod{\pi}}\) for all \({x \in \{A, B, 1, 2\}}\) if and only if fluxes are anti-symmetric modulo $\pi$, i.e.\ \({f_{A}^{+} = -f_{A}^{-} \mod{\pi}}\) and \({f_{B}^{+} = -f_{B}^{-} \mod{\pi}}\).

\subsection{Symmetric gauge in case of \texorpdfstring{$f_{A}^{+} + f_{B}^{+} = \pm(f_{A}^{-} + f_{B}^{-})$}{fAup + fBup = +-(fdownp + fBdown)}}
\label{sec:symmetric_gauge}

One might ask if there exists the gauge such that the phases \({\theta^{\pm}_{x}}\) are equal on all links $(x)$ in effective Hamiltonians~\eqref{eq:H_eff_Spin} and \eqref{eq:H_eff_Casrge}. In that case complex exponentials would factor out.
The answer is that if the total flux in $A$- and $B$-triangles, in other words total flux in square plaquette, satisfies
\begin{subequations}
\begin{align}
\label{eq:fA+fB=+}
	f_{A}^{+} + f_{B}^{+}
	&=
	f_{A}^{-} + f_{B}^{-}
\intertext{or}
\label{eq:fA+fB=-}
	f_{A}^{+} + f_{B}^{+}
	&=
	-\left(f_{A}^{-} + f_{B}^{-}\right)
,
\end{align}
\end{subequations}
then and only then there exists a gauge, which we refer to as the symmetric gauge, such that
\begin{subequations}
\begin{align}
\label{eq:theta_const_-}
	&\theta^{-}_{1}
	=
	\theta^{-}_{2}
	=
	\theta^{-}_{A}
	=
	\theta^{-}_{B}
	~\equiv~
	\theta^{-}
\intertext{or}
\label{eq:theta_const_+}
	&\theta^{+}_{1}
	=
	\theta^{+}_{2}
	=
	\theta^{+}_{A}
	=
	\theta^{+}_{B}
	~\equiv~
	\theta^{+}
\,,
\end{align}
\end{subequations}
respectively, where \({\theta^{\pm}_{x} = \varphi_{x}^{+} \pm \varphi_{x}^{-}}\).

Starting from the gauge with \({\varphi_{1}^{\pm} = \varphi_{2}^{\pm} = 0}\), it is straightforward to verify that if the condition~\eqref{eq:fA+fB=+} is fulfilled, the relation~\eqref{eq:theta_const_-} is achieved by choosing the gauge transformation parameters
\begin{subequations}
\begin{align}
	\alpha_{+} - \alpha_{-}
	&=
	-\frac{2}{3}
	\left(f_{A}^{+} - f_{A}^{-}\right)
	=
	\frac{2}{3}
	\left(f_{B}^{+} - f_{B}^{-}\right)
,
\\
	\beta_{+} - \beta_{-}
	&=
	\frac{1}{2}
	\left(\alpha_{+} - \alpha_{-}\right)
.
\end{align}
\end{subequations}
Here only the difference is significant, \({\alpha_{+} - \alpha_{-}}\) and \({\beta_{+} - \beta_{-}}\), while individual values of \({\alpha_{\sigma}}\) and \({\beta_\sigma}\) do not matter.

Similarly, if the condition~\eqref{eq:fA+fB=-} is fulfilled, the relation~\eqref{eq:theta_const_+} is achieved by choosing the gauge transformation parameters
\begin{subequations}
\begin{align}
	\alpha_{+} + \alpha_{-}
	&=
	-\frac{2}{3}
	\left(f_{A}^{+} + f_{A}^{-}\right)
	=
	\frac{2}{3}
	\left(f_{B}^{+} + f_{B}^{-}\right)
,
\\
	\beta_{+} + \beta_{-}
	&=
	\frac{1}{2}
	\left(\alpha_{+} + \alpha_{-}\right)
.
\end{align}
\end{subequations}


\begin{thebibliography}{77}
\setlength{\itemsep}{0pt}

\bibitem{AT_1943} J. Ashkin and E. Teller, 
\emph{Statistics of two-dimensional lattices with four components},
Phys. Rev. \textbf{64}, 178 (1943).

\bibitem{Baxter_1972} R. J. Baxter, 
\emph{Partition function of the eight-vertex lattice model},  
Ann. Phys. \textbf{70}, 193 (1972).

\bibitem{Baxter_Wu_1973} R. J. Baxter and F. Y. Wu, 
\emph{Exact xolution of an Ising model with three-spin interactions on a triangular lattice}, 
Phys. Rev. Lett. \textbf{31}, 1294 (1973).

\bibitem{Gross_Mezard_84} D. J. Gross and M. M\'ezard, 
\emph{The simplest spin glass}, 
Nucl. Phys. B \textbf{240}, 431 (1984).

\bibitem{Gardner_85} E. Gardner, 
\emph{Spin glasses with $p$-spin interactions}, 
Nucl. Phys. B \textbf{257}, 747 (1985).

\bibitem{m-spin_Ising_Kolb_Penson_86} M. Kolb and K. A. Penson, 
\emph{Conformal invariance and the phase transition of a spin chain with three-spin interaction}, 
J. Phys. A: Math. Gen. \textbf{19}, L779 (1986).

\bibitem{3-spin_1988_Penson} K. Penson, J. Debierre, and L. Turban, 
\emph{Conformal invariance and critical behavior of a quantum Hamiltonian with three-spin coupling in a longitudinal field},  
Phys. Rev. B \textbf{37}, 7884 (1988).

\bibitem{m-spin_Ising_1998} J. C. A. d'Auriac and F. Igl\'{o}i,  
\emph{Level statistics of multispin-coupling models with first- and second-order phase transitions},
Phys. Rev. E \textbf{58}, 241 (1998).

\bibitem{3-spin_Ising_1999} C. Tseng, S. Somaroo, Y. Sharf, E. Knill, R. Laflamme, T. F. Havel, and D. G. Cory, 
\emph{Quantum simulation of a three-body-interaction Hamiltonian on an NMR quantum computer}, 
Phys. Rev. A \textbf{61}, 012302 (1999).

\bibitem{3-spin_Ising_2009} X. Peng, J. Zhang, J. Du, and D. Suter, 
\emph{Quantum simulation of a system with competing two- and three-body interactions}, 
Phys. Rev. Lett. \textbf{103}, 140501 (2009).

\bibitem{Balents_Nature_2010} L. Balents, 
\emph{Spin liquids in frustrated magnets}, 
Nature \textbf{464}, 199 (2010). 

\bibitem{Anderson_RVB_87} P. W. Anderson, 
\emph{The resonating valence bond state in $\mathrm{La}_2\mathrm{CuO}_4$ and superconductivity}, 
Science \textbf{235}, 1196 (1987).

\bibitem{TP_w_FrcExc_1} D. S. Rokhsar and S. A. Kivelson, 
\emph{Superconductivity and the quantum hard-core dimer gas}, 
Phys. Rev. Lett. \textbf{61}, 2376 (1988).

\bibitem{TP_w_FrcExc_2} R. Moessner and S. L. Sondhi, 
\emph{Resonating valence bond phase in the triangular lattice quantum dimer model}, 
Phys. Rev. Lett. \textbf{86}, 1881 (2001).

\bibitem{TP_w_FrcExc_3} T. Senthil and M. P. A. Fisher, \emph{$Z_{2}$ gauge theory of electron fractionalization in strongly correlated systems}, 
Phys. Rev. B \textbf{62}, 7850 (2000).

\bibitem{TP_w_FrcExc_4} L. Balents, M. P. A. Fisher, and S. M. Girvin, 
\emph{Fractionalization in an easy-axis Kagome antiferromagnet}, 
Phys. Rev. B \textbf{65}, 224412 (2002).
 
\bibitem{TP_w_FrcExc_5} A. Kitaev, 
\emph{Anyons in an exactly solved model and beyond}, 
Ann. Phys. \textbf{321}, 2 (2006).

\bibitem{TP_w_FrcExc_6} P. A. Lee, N. Nagaosa, and X.-G. Wen, 
\emph{Doping a Mott insulator: Physics of high-temperature superconductivity}, 
Rev. Mod. Phys. \textbf{78}, 17 (2006).

\bibitem{Kalmeyer_Laughlin_87} V. Kalmeyer and R. B. Laughlin, 
\emph{Equivalence of the resonating-valence-bond and fractional quantum Hall states}, 
Phys. Rev. Lett. \textbf{59}, 2095 (1987).

\bibitem{WWZ_89} X.-G. Wen, F. Wilczek, and A. Zee, 
\emph{Chiral spin states and superconductivity}, 
Phys. Rev. B \textbf{39}, 11413 (1989).

\bibitem{Baskaran_89} G. Baskaran, 
\emph{Novel local symmetries and chiral-symmetry-broken phases in ${S=1/2}$ triangular-lattice Heisenberg model}, 
Phys. Rev. Lett. \textbf{63}, 2524 (1989).

\bibitem{Sen-Chjtra_95} D. Sen and R. Chitra, 
\emph{Large-$U$ limit of a Hubbard model in a magnetic field: Chiral spin interactions and paramagnetism}, 
Phys. Rev. B \textbf{51}, 1922 (1995).

\bibitem{Azbel_64} M. Ya Azbel', 
\emph{Energy spectrum of a conduction electron in a magnetic field}, 
Sov. Phys. JETP \textbf{19}, 634 (1964).
 
\bibitem{Brown_64} E. Brown, 
\emph{Bloch electrons in a uniform magnetic field}, 
Phys. Rev. \textbf{133}, A1038 (1964).

\bibitem{Hofstadter_76} R. Hofstadter, 
\emph{Energy levels and wave functions of Bloch electrons in rational and irrational magnetic fields}, 
Phys. Rev. B \textbf{14}, 2239 (1976).

\bibitem{Hasegava_etal_89} Y. Hasegawa, P. Lederer, T. M. Rice, and P. B. Wiegmann, 
\emph{Theory of electronic diamagnetism in two-dimensional lattices}, 
Phys. Rev. Lett. \textbf{63}, 907 (1989).

\bibitem{Barford_Kim_91} W. Barford and Ju H. Kim, 
\emph{Spinless fermions on frustrated lattices in a magnetic field}, 
Phys. Rev. B. \textbf{43}, 559 (1991).

\bibitem{Delibard_etal_RMP_2011} J. Dalibard, F. Gerbier, G. Juzeli\"{u}nas, and P. \"{O}hberg, 
\emph{Colloquium: Artificial gauge potentials for neutral atoms},
Rev. Mod. Phys. \textbf{83}, 1523 (2011).

\bibitem{Ketterle_etal_2013} H. Miyake, G. A. Siviloglou, C. J. Kennedy, W. C. Burton, and W. Ketterle, 
\emph{Realizing the Harper Hamiltonian with laser-assisted tunneling in optical lattices}, 
Phys. Rev. Lett. \textbf{111}, 185302 (2013).

\bibitem{Celi_etal_PRL_112_2014} A. Celi, P. Massignan, J. Ruseckas, N. Goldman, I. B. Spielman, G. Juzeli\"{u}nas, and M. Lewenstein, 
\emph{Synthetic gauge fields in synthetic dimensions}, 
Phys. Rev. Lett. \textbf{112}, 043001 (2014).

\bibitem{Galitski_PT_2019} V. Galitski, G. Juzeli\"{u}nas, and I. B. Spielman, 
\emph{Artificial gauge fields with ultracold atoms}, 
Phys. Today \textbf{72}, 38 (2019).

\bibitem{UCA_triangle_2010} C. Becker, P. Soltan-Panahi, J. Kronj\"{a}ger, S. D\"{o}rscher, K. Bongs, and K. Sengstock, 
\emph{Ultracold quantum gases in triangular optical lattices},  
New J. Phys. \textbf{12}, 065025 (2010).

\bibitem{Triangular_lattice_2019} V. T. Phong, Z. Addison, S. Ahn, H. Min, R. Agarwal, and E. J. Mele, 
\emph{Optically controlled orbitronics on a triangular lattice}, 
Phys. Rev. Lett. \textbf{123}, 236403 (2019).

\bibitem{UCA_Kagome_Lattice_2012} G.-B. Jo, J. Guzman, C. K. Thomas, P. Hosur, A. Vishwanath, and D. M. Stamper-Kurn, 
\emph{Ultracold atoms in a tunable optical Kagome lattice}, 
Phys. Rev. Lett. \textbf{108}, 045305 (2012).

\bibitem{Tarruel_12} L. Tarruell, D. Greif, T. Uehlinger, G. Jotzu, and T. Esslinger, 
\emph{Creating, moving and merging Dirac points with a Fermi gas in a tunable honeycomb lattice}, 
Nature \textbf{483}, 302 (2012).

\bibitem{Esslinger_13} T. Uehlinger, G. Jotzu, M. Messer, D. Greif, W. Hofstetter, U. Bissbort, and T. Esslinger, 
\emph{Artificial graphene with tunable interactions}, 
Phys. Rev. Lett. \textbf{111}, 185307 (2013).

\bibitem{Zig-Zag_ladder_2016} E. Anisimovas, M. Ra\v{c}i\={u}nas, C. Str\"{a}ter, A. Eckardt, I. B. Spielman, and G. Juzeli\={u}nas, 
\emph{Semisynthetic zigzag optical lattice for ultracold bosons}, 
Phys. Rev. A \textbf{94}, 063632 (2016). 

\bibitem{Kang_etal_18} J. H. Kang, J. H. Han, and Y. Shin, 
\emph{Realization of a cross-linked chiral ladder with neutral fermions in a 1D optical lattice by orbital-momentum coupling}, 
Phys. Rev. Lett. \textbf{121}, 150403 (2018).

\bibitem{Kang_etal_20} J. H. Kang, J. H. Han, and Y. Shin, 
\emph{Creutz ladder in a resonantly shaken 1D optical
lattice}, 
New J. Phys. \textbf{22}, 013023 (2020).

\bibitem{Taglieber_etal_08} M. Taglieber, A.-C. Voigt, T. Aoki, T.W. H{\"a}nsch, and K. Dieckmann, 
\emph{Quantum degenerate two-species Fermi-Fermi mixture coexisting with a Bose-Einstein condensate}, 
Phys. Rev. Lett. \textbf{100}, 010401 (2008).

\bibitem{Orignac_Giamarchi_01} E. Orignac and T. Giamarchi, 
\emph{Meissner effect in a bosonic ladder}, 
Phys. Rev. B \textbf{64}, 144515 (2001).

\bibitem{Bloch_etal_2014} M. Atala, M. Aidelsburger, M. Lohse, J. T. Barreiro, B. Paredes, and I. Bloch, 
\emph{Observation of chiral currents with ultracold atoms in bosonic ladders}, 
Nature Phys. \textbf{10}, 588 (2014).

\bibitem{Mazza_etal_2016} S. Barbarino, L. Taddia, D. Rossini, L. Mazza, and R. Fazio, 
\emph{Synthetic gauge fields in synthetic dimensions: interactions and chiral edge modes}, 
New J. Phys. \textbf{18}, 035010 (2016).

\bibitem{Zoller_etal-2017} J. C. Budich, A. Elben, M. Lacki, A. Sterdyniak, M. A.Baranov, and P. Zoller, 
\emph{Coupled atomic wires in a synthetic magnetic field}, 
Phys. Rev. A \textbf{95}, 043632 (2017).

\bibitem{Dalmonte_etal_2017} M. Calvanese Strinati, E. Cornfeld, D. Rossini, S. Barbarino, M. Dalmonte, R. Fazio, E. Sela, and L. Mazza, 
\emph{Laughlin-like states in Bosonic and Fermionic atomic synthetic ladders}, 
Phys. Rev. X \textbf{7}, 021033 (2017).

\bibitem{Dalmonte_etal_2018} S. Barbarino, M. Dalmonte, R. Fazio, and G. E. Santoro, 
\emph{Topological phases in frustrated synthetic ladders with an odd number of legs}, 
Phys. Rev. A \textbf{97}, 013634 (2018).

\bibitem{Lewenstein_etal_17} J. J\"{u}nemann, A. Piga, S.-J. Ran, M. Lewenstein, M. Rizzi, and A. Bermudez, 
\emph{Exploring interacting topological insulators with ultracold atoms: The synthetic Creutz-Hubbard model}, 
Phys. Rev. X \textbf{7}, 031057 (2017).

\bibitem{Lewenstein_etal_19} E. Tirrito, M. Rizzi, G. Sierra, M. Lewenstein, and A. Bermudez, 
\emph{Renormalization group flows for Wilson-Hubbard matter and the topological Hamiltonian}, 
Phys. Rev. B \textbf{99}, 125106 (2019).
  
\bibitem{Lewenstein_etal_22} E. Tirrito, M. Lewenstein, and A. Bermudez, 
\emph{Topological chiral currents in the Gross-Neveu model extension}, 
Phys. Rev. B \textbf{106}, 045147 (2022).

\bibitem{spin_flux} M. Aidelsburger, M. Atala, M. Lohse, J. T. Barreiro, B. Paredes, and I. Bloch, 
\emph{Realization of the Hofstadter Hamiltonian with ultracold atoms in optical lattices}, 
Phys. Rev. Lett. \textbf{111}, 185301 (2013).

\bibitem{spin_gauge_1} T. Fujita, M. B. A. Jalil, S. G. Tan, and S. Murakami, 
\emph{Gauge fields in spintronics}, 
J. Appl. Phys. \textbf{110}, 121301 (2011).

\bibitem{spin_gauge_2} G. Tatara, 
\emph{Effective gauge field theory of spintronics}, 
Physica E \textbf{106}, 208 (2019).

\bibitem{spin_book} S. Maekawa, S. O. Valenzuela, E. Saitoh, and T. Kimura, 
\emph{Spin Current}, 
Oxford University Press (2012).

\bibitem{NCN_05} B. N. Narozhny, S. T. Carr, and A. A. Nersesyan, 
\emph{Fractional charge excitations in fermionic ladders}, 
Phys. Rev. B \textbf{71}, 161101 (2005).

\bibitem{CNN_06} S. T. Carr, B. N. Narozhny, and A. A. Nersesyan, 
\emph{Spinless fermionic ladders in a magnetic field: Phase diagram}, 
Phys. Rev. B \textbf{73}, 195114 (2006).

\bibitem{BKT_transition} V. L. Berezinskii, 
\emph{Destruction of long--range order in one--dimensional and two--dimensional Systems possessing a continuous symmetry group. II. Quantum systems}, 
Sov. Phys. JETP \textbf{34}, 610 (1972);    
J. M. Kosterlitz and D. J. Thouless, 
\emph{Ordering, metastability and phase transitions in two-dimensional systems}, 
J. Phys. C: Solid State Phys. \textbf{6}, 1181 (1973).

\bibitem{Cazalilla_etal_2022} C.-H. Huang, M. Tezuka, and M. A. Cazalilla, 
\emph{Topological Lifshitz transitions, orbital currents, and interactions in low-dimensional Fermi gases in synthetic gauge fields}, 
New J. Physics \textbf{24}, 033043 (2022).

\bibitem{Dalmonte_etal_2023} Z. Bacciconi, G. M. Andolina, T. Chanda, G. Chiriac\'o, M. Schir\'o, and M. Dalmonte, 
\emph{First-order photon condensation in magnetic cavities: A two-leg ladder model}, 
SciPost Phys. \textbf{15}, 113 (2023).

\bibitem{BN_1_EPJB} B. Beradze and A. Nersesyan, 
\emph{Spectrum, Lifshitz transitions and orbital current in frustrated fermionic ladders with a uniform flux}, 
Eur. Phys. J. B \textbf{96} 2 (2023).

\bibitem{BN_2_PRB} B. Beradze, M. Tsitsishvili, E. Tirrito, M. Dalmonte, T. Chanda, A. Nersesyan, 
\emph{Emergence of non-Abelian $SU(2)$ invariance in Abelian frustrated fermionic ladders}, 
Phys. Rev. B \textbf{108}, 075146 (2023)

\bibitem{Giamarchi_23} C.-M. Halati and T. Giamarchi, 
\emph{Bose-Hubbard triangular ladder in an artificial gauge field}, 
Phys. Rev. Res. \textbf{5}, 013126 (2023).

\bibitem{Bulaevskii_etal_07} L. N. Bulaevskii, C. D. Batista, M. Mostovoy, and D. Khomskii, 
\emph{Electronic orbital currents and polarization in Mott insulators}, 
Phys. Rev. B \textbf{78}, 024402 (2008).

\bibitem{Bulaevskii_etal_09} K. A. Al-Hassanieh, C. D. Batista, G. Ortiz, and L. N. Bulaevskii, 
\emph{Field-induced orbital antiferromagnetism in Mott insulators}, 
Phys. Rev. Lett., \textbf{103}, 216402 (2009).

\bibitem{Anderson59} P. W. Anderson, 
\emph{New approach to the theory of superexchange interactions}, 
Phys. Rev. \textbf{115}, 2 (1959).

\bibitem{Bulaevskii67} L. N. Bulaevskii, 
\emph{Quasihomopolar electron levels in crystals and molecules}, 
Sov. Phys. JETP \textbf{24}, 154 (1967).

\bibitem{Takahashi77} M. Takahashi, 
\emph{Half-filled Hubbard model at low temperature}, 
J. Phys. C: Solid State Phys. \textbf{10}, 1289 (1977).
                    
\bibitem{Harris_Lange_67} A. B. Harris and R. V. Lange, 
\emph{Single-particle excitations in narrow energy bands}, 
Phys. Rev. \textbf{157}, 295 (1967).

\bibitem{MacDonald88} A. H. MacDonald, S. M. Girvin, and D. Yoshioka, 
\emph{$t/U$ expansion for the Hubbard model}, 
Phys. Rev. B \textbf{37}, 9753 (1988).
                  
\bibitem{GMJ_2016} I. Grusha, M. Menteshashivi, and G. I. Japaridze, 
\emph{Effective Hamiltonian for a half-filled asymmetric ionic Hubbard chain with alternating on-site interaction}, 
Int. J. Mod. Phys. B \textbf{30}, 1550260 (2016).

\bibitem{SW_66} J. R. Schrieffer and P. A. Wolff, 
\emph{Relation between the Anderson and Kondo Hamiltonians}, 
Phys. Rev. \textbf{149}, 491 (1966).
                    
\bibitem{Hubbard-Operators-Book} S. G. Ovchinnikov, V. V. Val'kov, 
\emph{Hubbard operators in the theory of strongly correlated electrons}, 
Imperial College Press (2004).

\bibitem{Micnas_etal_90} R. Micnas, J. Ranninger, and S. Robaszkiewicz, 
\emph{Superconductivity in narrow-band systems with local nonretarded attractive interactions}, 
Rev. Mod. Phys. \textbf{62}, 113 (1990).

\bibitem{Frahm_1997} H. Frahm and C. R\"{o}denbeck, 
\emph{Properties of the chiral spin liquid state in generalized spin ladders}, 
J. Phys. A: Math. Gen. \textbf{30} 4467 (1997).

\bibitem{Schmoll_2019} P. Schmoll, A. Haller, M. Rizzi, and R. Or\'{u}s, 
\emph{Quantum criticality on a chiral ladder: An SU(2) infinite density matrix renormalization group study}, 
Phys. Rev. B \textbf{99}, 205121 (2019).

\bibitem{FK_69} L. M. Falicov and J. C. Kimball, 
\emph{Simple model for semiconductor-metal transitions: $\mathrm{SmB}_6$ and transition-metal oxides}, 
Phys. Rev. Lett. \textbf{22}, 997 (1969).

\bibitem{ising_3spin} E. Jur\v{c}i\v{s}inov\'{a} and M. Jur\v{c}i\v{s}in, 
\emph{Solution of the antiferromagnetic Ising model with multisite interaction on a zigzag ladder}, 
Phys. Rev. E \textbf{90}, 032108 (2014).

\bibitem{1D_Hubbard_book} F. H. L. Essler, H. Frahm, F. G\"ohmann, A. Klümper, V. E. Korepin, 
\emph{The one-dimensional Hubbard model}, 
Cambridge University Press (2010).


\end{thebibliography}
\end{document}